\documentclass[]{interact}
\usepackage{graphicx}
\usepackage{dcolumn}
\usepackage{bm}
\usepackage{epstopdf}
\usepackage[caption=false]{subfig}

\usepackage[numbers,sort&compress,merge]{natbib}

\bibpunct[, ]{[}{]}{,}{n}{,}{,}

\theoremstyle{plain}

\theoremstyle{definition}

\theoremstyle{remark}

\begin{document}

\title{Rayleigh-Brillouin light scattering spectroscopy of air; experiment, predictive model and dimensionless scaling}

\author{
\name{Yuanqing Wang\textsuperscript{a},
Ziyu Gu\textsuperscript{a},
Kun Liang\textsuperscript{b} and Wim Ubachs\textsuperscript{a}\thanks{CONTACT Wim Ubachs. Email: w.m.g.ubachs@vu.nl}}
\affil{\textsuperscript{a}Department of Physics and Astronomy, LaserLaB, Vrije Universiteit,  De Boelelaan 1081, 1081 HV Amsterdam, The Netherlands;}
\textsuperscript{b}School of electronic information and communications, Huazhong University of Science and Technology, Wuhan 430074, China
}

\date{\today}
\maketitle
\begin{abstract}
Spontaneous Rayleigh-Brillouin scattering (RBS) experiments have been performed in air for pressures in the range 0.25 - 3 bar and temperatures in the range 273 - 333 K. The  functional behaviour of the RB-spectral profile as a function of experimental parameters, such as the incident wavelength, scattering angle, pressure and temperature is analyzed, as well as the dependence on thermodynamic properties of the gas, as the shear viscosity, the thermal conductivity, the internal heat capacity and the bulk viscosity. Measurements are performed in a scattering geometry detecting at a scattering angle $\theta=55.7^\circ$ and an incident wavelength of $\lambda_i=532.22$ nm, at which the Brillouin features become more pronounced than in a right angles geometry and for ultraviolet light. For pressure conditions of 1 - 3 bar the RB-spectra, measured at high signal-to-noise ratio, are compared to Tenti-S6 model calculations and values for the bulk viscosity of air are extracted. Values of $\eta_b$ are found to exhibit a linear dependence on temperature over the measurement interval in the range $1.0 - 2.0 \times 10^{-5}$ Pa$\cdot$s.
A temperature dependent value is deduced from a collection of experiments to yield: $\eta_{\rm b} = (0.86  \times 10^{-5}) + 1.29 \times 10^{-7} \cdot (T - 250)$.
These results are implemented in model calculations that were verified for the low pressure conditions ($p < 1$ bar) relevant for the Earth's atmosphere. As a result we demonstrate that the RB-scattering spectral profiles for air under sub-atmospheric conditions can be generated via the Tenti-S6 model, for given gas-phase and detection conditions ($p$, $T$, $\lambda_i$, and $\theta$), and for values for the gas transport coefficients.
Spectral profiles for coherent RB-scattering in air are also computed, based on the Tenti-S6 formalism, and the predictions are compared with profiles of spontaneous RBS.
Finally data on RB-scattering in air, obtained under a variety of pressure, temperature, wavelength and scattering angles, are analyzed in terms of universal scaling, involving the dimensionless uniformity parameter $y$ and the dimensionless frequency $x$. Such scaling behaviour is shown to be well behaved for a wide parameter space and implies that RB-scattering spectra can be generated for a wide range of atmospheric applications of RB-scattering. The verification of this dimensionless scaling also shows that air can be treated as an ideal gas in the atmospheric regime, where $y \leq 1$.
\end{abstract}

\begin{keywords}
Spontaneous Rayleigh-Brillouin scattering; Light scattering; Earth's atmosphere; Gas transport coefficients; Ideal gases
\end{keywords}

\section{\label{sec:intro}Introduction}

Light scattering phenomena and the measurement of spectral profiles of Rayleigh-Brillouin (RB) scattered light is of great relevance in a variety of applications~\cite{Miles2001}. Foremost, studies of the Earth's atmosphere are often based on the analysis of light scattering, e.g. via satellite remote sensing, where the Doppler-wind lidar of the ADM-Aeolus mission is a marked example~\cite{Stoffelen2006,Reitebuch2009,Marksteiner2018,Savli2019,Lux2020,Witschas2020}. Temperature profiles in the atmosphere can be probed from RB-scattered light in Lidar configurations~\cite{Witschas2014b}.
In industrial environments and in combustion light scattering is applied for thermometry~\cite{Sutton2004}, for probing gas phase velocity fields~\cite{Forkey1998,Boguszko2005,Doll2017}, complex flows~\cite{Miles1997} and jets~\cite{Benhassen2017}.

Information on the measurement conditions, such as the density and temperature of the gas, but also on the intrinsic properties of the gas, such as the mass of the molecules, the thermal conductivity, the viscosity and the internal heat capacity is contained in the spectral line shape of the RB-scattered light. That means that some of these parameters can be extracted from dedicated light scattering experiments. So can the temperature of air be determined from the RB-line shape~\cite{Witschas2014}, but so can also the viscosity of gases be determined~\cite{Gu2013b,Wang2019}.

The overall behaviour of RB-spectra can be cast into a generalized description to depend on the uniformity parameter $y$,
a dimensionless parameter dividing the physics of light scattering of a gas into different regimes.
For $y \ll 1$, the scattering is in the Knudsen regime and governed by the Doppler effect, for which the profile can be expressed by a Gaussian function.
For $y \gg 1$ the domain of hydrodynamics is entered where a gas is treated as a continuum fluid where a threefold resonance structure is obtained with two fully distinct Brillouin side-peaks and the spectrum represented by three Lorentzian functions~\cite{Mountain1966}. The hydrodynamic approach can be extrapolated to cover the regime of somewhat lower pressures~\cite{Hammond1976}.
In the intermediate regime, of $y \sim 1$, Brillouin side peaks marginally overlap with the central Rayleigh peak, making the profile complex and no exact solutions are available.

Various models have been developed to describe the light scattering spectral line shape in this intermediate, or kinetic regime, which is of relevance for the Earth's atmosphere.
Models are expressed in terms of gas transport coefficients, and are generally based on linearization or approximate treatment of the Boltzmann equation representing the collisions in the medium. One approach is the Tenti-model~\cite{Boley1972,Tenti1974} providing a rather straightforward representation of the scattering profile in terms of six moments connected to transport coefficients.
The Tenti approach has been shown to work well in the case of diatomic and linear molecules~\cite{Gu2013b,Gu2014a,Wang2018,Shang2019}, where usually the bulk viscosity is included in the model as a fit parameter. Another approach for the kinetic regime dealt with collisions of molecules treated as rough spheres, which well described the RB-spectrum of SF$_6$~\cite{Wang2017}.

Laboratory experiments on RB-scattering of gases typically serve the purpose of verifying and optimizing the models and determine the thermodynamic gas transport coefficients that are essential building blocks for models of the light scattering spectral profiles. For this reason a number of RB-scattering studies were performed on air, in view of its relevance for atmospheric science. RB-profiles of air were determined over a variety of pressures and temperatures as well as incident wavelengths (366 nm and 403 nm), where a comparison was made with scattering in pure nitrogen and pure oxygen, while a model description of the Tenti-model was pursued~\cite{Witschas2010,Witschas2011,Witschas2014,Gu2014b}.
\citet{Shang2019b} measured the spectra of air at an incident wavelength of 532 nm at elevated pressures of 4.0 - 7.0 bar for varying temperatures, thus extending the parameter space of RB-scattering studies.

Most of the laboratory studies were performed under scattering angles of 90$^\circ$. In the present study the parameter space for RB-scattering in air at 532 nm is widened to the study at a sharper scattering angle, in the forward direction, which bears the advantage that Brillouin side peaks become more pronounced. Also pressure and temperature were varied in the measurements of spectral profiles. The study verifies the validity of the Tenti description for RB scattering in air, and results in values for the bulk viscosity of air.

\section{The Rayleigh-Brillouin lineshape}
\label{simulations}

\begin{figure}[b]
  \centering
  \includegraphics[width=0.5\linewidth]{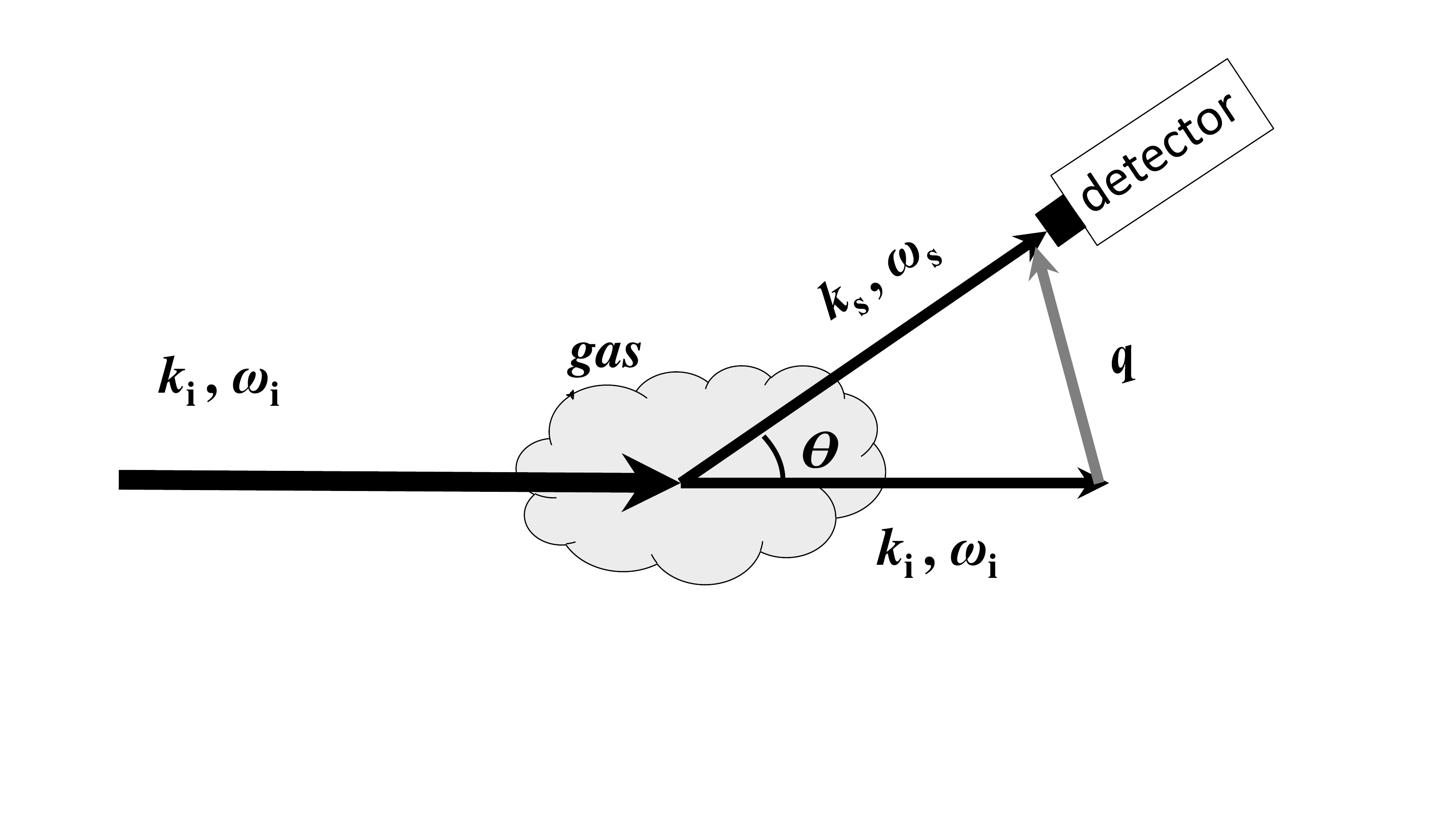}
  \caption{The scattering vector diagram for the detection of Rayleigh scattering. $\boldsymbol{k_{\rm i}}$, $\boldsymbol{k_{\rm s}}$, $\omega_{\rm i}$, $\omega_{\rm s}$ are the incident and scattered light wave vectors and frequencies respectively. $\boldsymbol{q}= \boldsymbol{k_{\rm s}}-\boldsymbol{k_{\rm i}}$ is the scattering wave vector.}
  \label{Fig:Scattervector}
\end{figure}

At low pressures in the gas phase, in the regime of mbars or the Knudsen regime, collective effects are absent, light scattering is elastic and can be well approximated as pure Rayleigh scattering. The elastic peak depends on the angle under which the scattered radiation is detected, as illustrated in Fig.~\ref{Fig:Scattervector}. Since in light scattering experiments the magnitudes of scattered and incident wave vectors are similar, $k_s \approx k_i = 2\pi n/\lambda_i$, with $n$ the refractive index and $\lambda_i$ the incident wavelength, the magnitude of the scattering wave vector is represented by:
\begin{equation}
  q = 2k_i \sin{\frac{\theta}{2}} = \frac{4\pi n}{\lambda_i} \sin{\frac{\theta}{2}}  \ .
  \label{q-vec}
\end{equation}

\begin{figure}
  \centering
  \includegraphics[width=0.5\linewidth]{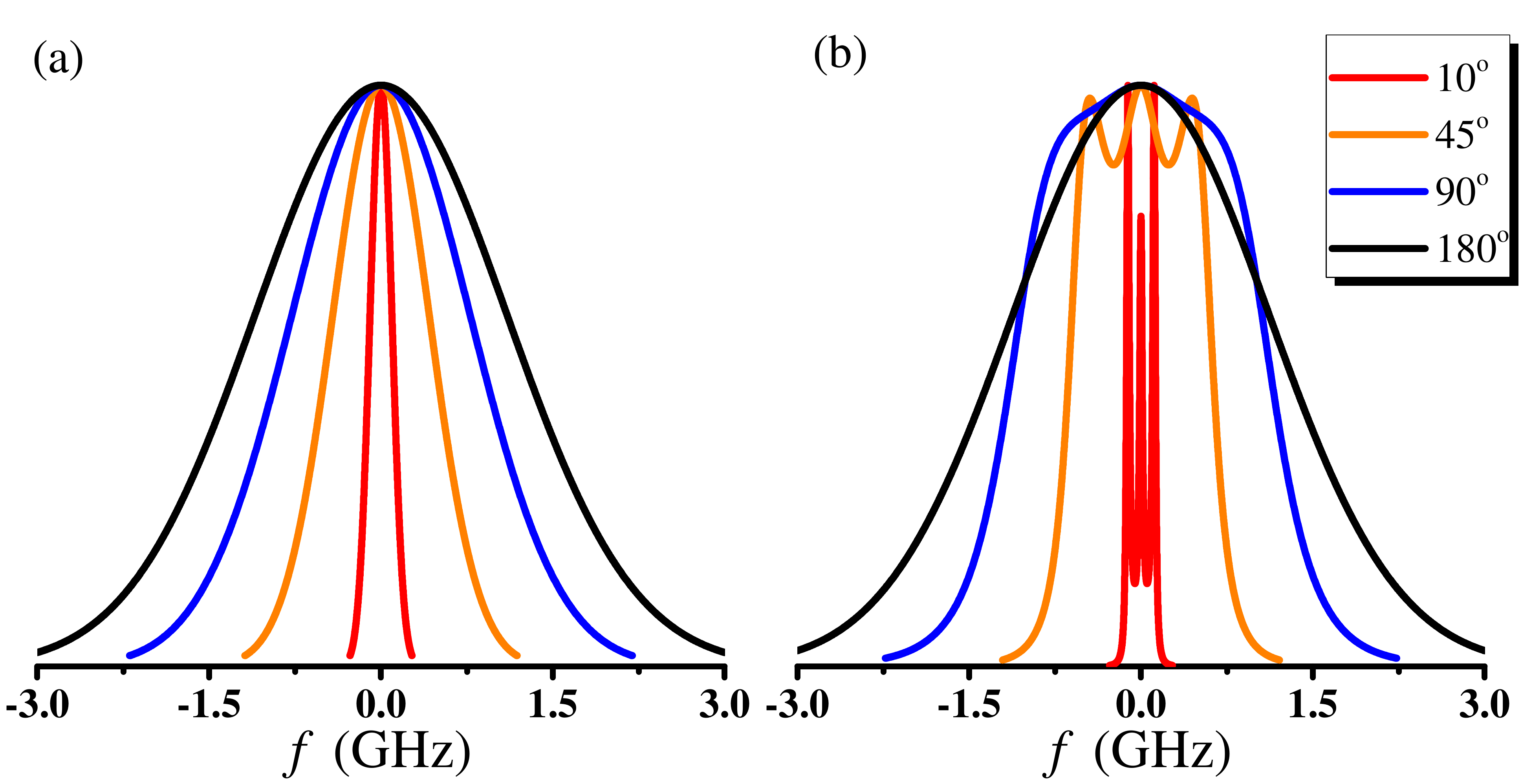}
  \caption{The angle dependence ($\theta$) of the light scattering spectral profile of air simulated by the Tenti-S6 model. (a) Rayleigh scattering in the Knudsen regime; (b) Rayleigh-Brillouin scattering at 1 bar pressure. Further parameters: $T=20\,^\circ$C and $\lambda_i=532$ nm.}
  \label{Fig:Angle}
\end{figure}

Rayleigh scattering of a gas involves the thermal motion of the molecules. Depending on the velocity $v$ the detected photon frequency is shifted from the emitted central frequency $f_0$ via:
\begin{equation}
  f = f_0 + v\frac{q}{2\pi} =  f_0 + 2n f_0 \frac{v}{c} \sin{\frac{\theta}{2}} \ ,
\label{v-scat}
\end{equation}
where $c$ is speed of light in vacuum. Under thermal conditions the molecules in a gas exhibit a velocity distribution corresponding to the Maxwell distribution:
\begin{equation}
    \phi_M(v) = (\pi v_0^2)^{-3/2} {\rm e}^{-v^2/v_0^2}
\end{equation} with $v_0$ the thermal velocity  ($v_0=\sqrt{2k_BT/m}$), $k_B$ the Boltzmann constant, $T$ the temperature, and $m$ the molecular mass. Insertion of this velocity distribution function into Eq.~\eqref{v-scat} results in a spectral function centered at $f_0$, with a Gaussian functional form of width:
\begin{equation}
  \Delta f_D =   \frac{2n f_0}{c} \sin{\frac{\theta}{2}} \sqrt{\frac{2\ln2 k_B T}{m}}  \ .
\label{Ray-Dopp}
\end{equation}
This is the Doppler effect associated with the molecular motion, turning the elastic Rayleigh scattering peak into a spectral profile of Gaussian nature. While the Doppler width depends on the incident wavelength, the molecular mass, and the temperature, it also strongly depends on the scattering angle $\theta$, as illustrated in Fig.~\ref{Fig:Angle}(a). At smaller scattering angles the scattering profile becomes narrower, and for scattering in the exact forward direction only a coherent wave remains~\cite{Miles2001}. This overall width of the scattering profile is retained under conditions of collisions in the kinetic regime.

At somewhat higher pressures, including the regime of the Earth's atmosphere at various altitudes, collisions set in to play a role in the light scattering process, giving rise to a fine structure of the Rayleigh line, predicted by Brillouin~\cite{Brillouin1922}, and independently by Mandel'shtam~\cite{Mandelstam1926}, which will be referred to in the following as the Rayleigh-Brillouin lineshape.
Fluctuations in the pressure or concentration of the medium are the source of elastic waves that propagate in all directions. By taking a fixed geometry with an incident vector $\boldsymbol{k_{\rm i}}$
and detection of a scattered vector $\boldsymbol{k_{\rm s}}$ in the direction of angle $\theta$, scattering vectors $\boldsymbol{q}$ associated with certain elastic waves fulfilling the Bragg condition will be selected for maximum scattering intensity~\cite{Fabelinskii2012}. The magnitude $q$ of the scattering vector is similar to that given in Eq.~(\ref{q-vec}). This form of scattering can also be considered as a Doppler effect, where the change in light frequency is not a result of scattering from a molecule at velocity $v$, but rather a reflection from a traveling elastic wave propagating through the medium at the sound velocity $v_s$. This scattering process gives rise to positive and negative frequency shifts $\pm \Omega_B$ in units of angular frequency:
\begin{equation}
    \Omega_B = \frac{4 \pi n v_s}{\lambda} \sin \frac{\theta}{2}.
    \label{Bril-shift}
\end{equation}
Hence the Brillouin shift scales with the angle of observation as shown in Fig.~\ref{Fig:Angle}.

Under such conditions of increased density or pressure the RB-light scattering spectral profile is determined not only by ($T$, $\lambda_i$, $m$, $\theta$), but by additional thermodynamic properties of the gas, such as the pressure $p$, the thermal conductivity $\kappa$, the shear viscosity $\eta_{\rm s}$, the internal specific heat capacity $c_{\rm int}$, and a multitude of relaxation processes occurring in the gas, usually expressed in terms of a bulk viscosity $\eta_{\rm b}$. The latter are known as the macroscopic gas transport coefficients. Air is treated as consisting of a hypothetical single component gas of diatomic molecules of mass $m=29.0$ amu, the geometric mean of the masses of nitrogen and oxygen accounting for their abundances, with its experimental gas transport parameters~\cite{Gu2014b}.  So the overall width of the scattering feature, mainly determined by $\Delta f_D$, depending on the molecular mass $m$, is rather similar. It is noted that the minor constituents of air do not have a significant influence on the RB-scattering spectrum of air, as was experimentally verified for water vapour saturated air samples~\cite{Witschas2010}.

\begin{figure*}
  \centering
  \includegraphics[width=0.75\linewidth]{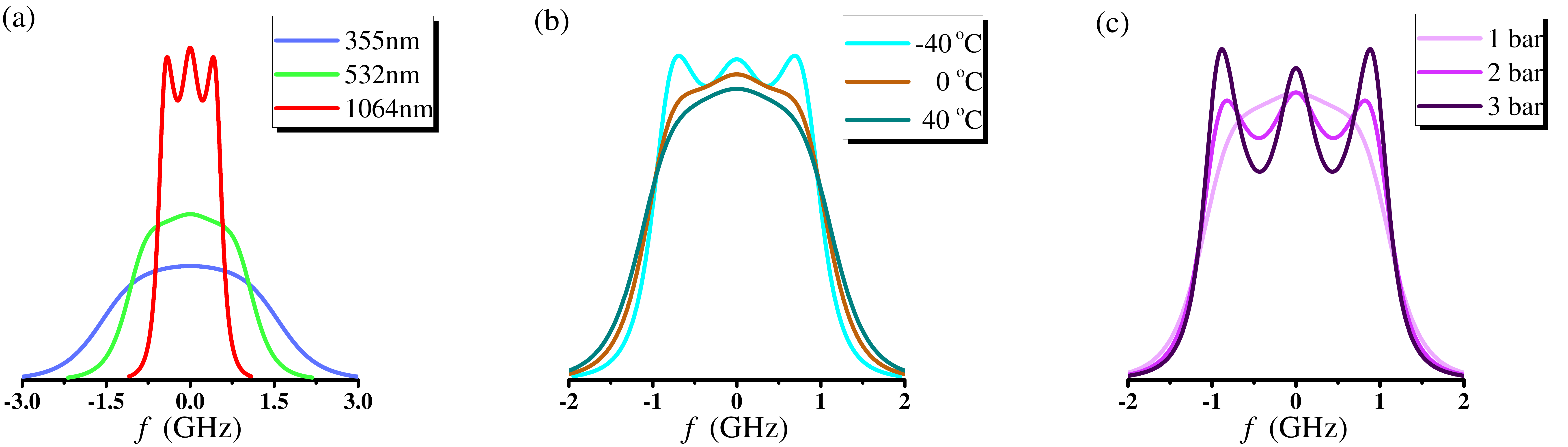}
  \caption{Simulated spectral line shapes for spontaneous RB-scattering in air by the Tenti-S6 model plotted on a scale of normalized integrated intensity. (a) Wavelength dependence; (b) Temperature dependence; (c) Pressure dependence.  Generally parameters are $T=20\,^{\circ}$C, $\lambda_i=532$ nm, $p=1$ bar, $\theta=90^\circ$, unless otherwise specified. }
  \label{Fig:RB-Lineshape}
\end{figure*}

Collisions in gaseous media are treated by the Boltzmann kinetic equation, which was extended to the case of gases with internal degrees of freedom by Wang Chang and Uhlenbeck~\cite{Wang1951} to the full WCU kinetic equation for the phase-space distribution function $f(\boldsymbol r,\boldsymbol v,t)$:
\begin{eqnarray}
    \frac{\partial f_i(\boldsymbol r,\boldsymbol v,t)}{\partial t} + \boldsymbol v \cdot \nabla f_i(\boldsymbol r,\boldsymbol v,t) = \\ \nonumber
    \sum_{jkl} \int |\boldsymbol v- \boldsymbol v_1| \sigma^{kl}_{ij} (f_k'f_l' - f_if_j) d\Omega d^3v_1 \ ,
    \label{Eq:FullWCUFunc}
\end{eqnarray}
where $\boldsymbol{r}$ is the space position and $\boldsymbol{v}$ is the velocity of the particle at time $t$. Indexes $i$, $j$, $k$ and $l$ are internal states of the molecules ($v$ and $v_1$ velocities of first and second molecules) exhibiting elastic and inelastic collisions via state-to-state cross sections $\sigma$. As collisions between molecules only cause small deviations from equilibrium, the distribution function of the $i^{th}$ internal level can be linearized via:
\begin{equation}
    f_i(\boldsymbol r,\boldsymbol v,t) = n_0 x_i \phi_M(v) [1+ h_i(\boldsymbol r,\boldsymbol v,t)] \ ,
    \label{Eq:DistributionFunc}
\end{equation}
with $n_0$ the number density, $\phi_M(v)$ the Maxwell distribution function, $x_i$ the fraction of molecules in state $i$ and $h_i(\boldsymbol r,\boldsymbol v,t)$ the deviation from equilibrium.

The thermodynamic properties of a gas can then be expressed in a perturbative manner, as deviations from equilibrium in terms of $h$. Boley et al~\cite{Boley1972} developed a model, based on the WCU-equation, involving non-degenerate eigenvectors, that were written as such linearized approximations.
These seven eigenvectors are related to seven physical moments (1) the fraction of particles in different internal states, (2) the momentum, (3) the translational energy, (4) the translational heat flux, (5) the internal energy, (6) the internal heat flux and (7) the traceless pressure tensor.
This model, later referred to as the Tenti-S7 model, was subsequently truncated into a 6-dimensional model (Tenti-S6) by Tenti et al.~\cite{Tenti1974} by neglecting this seventh parameter. The truncated S6-version of the Tenti-model is considered to be the standard for treating RBS profiles in the kinetic regime. The description in terms of six independent eigenvectors and some detail of the mathematical evaluation is presented in the Supplementary Material, and also a numerical model in the form of a Matlab-code is provided in the Supplementary Material.

With these approximations the WCU-equation can be simplified into a vectorial equation:
\begin{equation}
    \frac{\partial h}{\partial t} + \boldsymbol v \cdot \nabla h = n_0 \boldsymbol J h
    \label{Eq:LinearBoltzmanTentiS6} \ ,
\end{equation}
where $\boldsymbol J$ is the collision operator, in the form of a $N \times N$ matrix (of dimension 6 for the Tenti-S6 model), and each matrix element represents a collision integral. This equation is used to generate a linear system of equations, by expanding $h$ into eigenfunctions of $\boldsymbol J$.
The six eigenvectors of the Tenti-S6 model can be expressed in terms of three transport coefficients (the shear viscosity $\eta_{\rm s}$, the bulk viscosity $\eta_{\rm b}$ and the thermal conductivity $\kappa$), the atomic mass $m$ of the particles and the internal specific heat capacity per molecule c$_{\rm int}$.
In this way the model calculations are intimately connected to well-known and measurable thermodynamic properties of the gas.

\begin{figure*}
  \centering
  \includegraphics[width=1.0\linewidth]{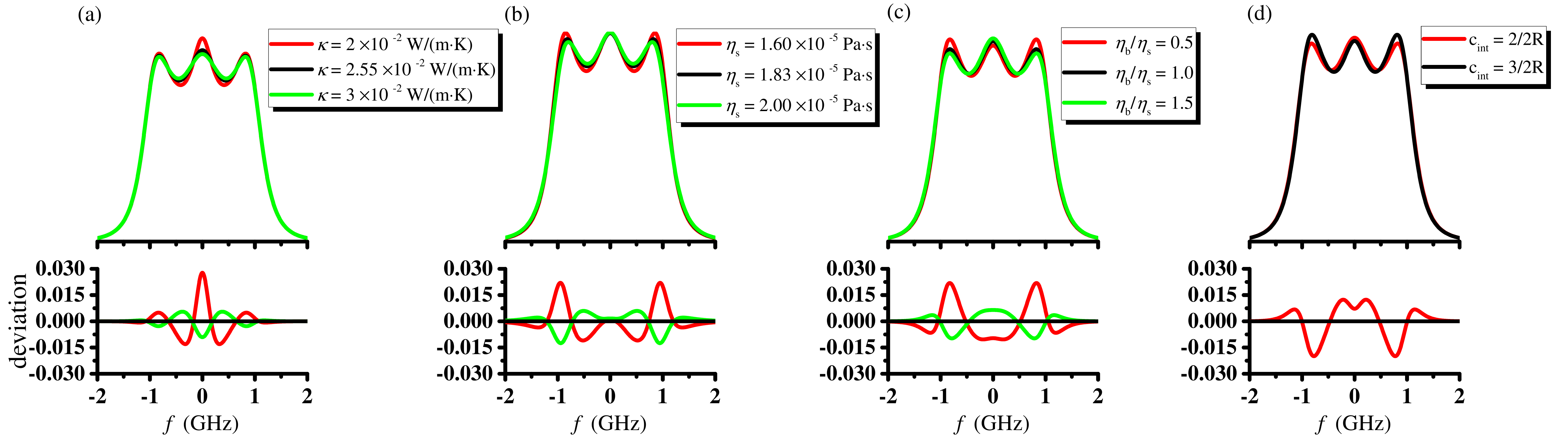}
  \caption{Simulated spectral line shapes for spontaneous RB-scattering in air by the Tenti-S6 model depending on various transport coefficients. The spectra are plotted on a scale of normalized integrated intensity, while below the graphs residuals are presented with respect to the central value of the parameter. (a) Thermal conductivity $\kappa$; (b) Shear viscosity $\eta_{\rm s}$; (c)  Ratio of bulk viscosity vs. shear viscosity $\eta_{\rm b}/\eta_{\rm s}$; (d) Internal heat capacity $c_{\rm int}$.  Parameters $T=20\,^{\circ}$C, $\lambda_i=532$ nm, $\theta=90^\circ$, $p=2$ bar are fixed. Values for transport coefficients are $\kappa$ = 2.55$\times 10^{-2}$ W/m$\cdot$K, $\eta_{\rm s}$ = 1.83$\times10^{-5}$ Pa$\cdot$s, $\eta_{\rm b}$ = 1.48$\times10^{-5}$ Pa$\cdot$s, $c_{\rm int}$ =  2/2R, unless varied. See further main text.
  }
  \label{Fig:RB-Transport}
\end{figure*}

Based on this framework Tenti-model codes were developed to straightforwardly calculate the spectral line shape of Rayleigh-Brillouin scattering~\cite{Boley1972,Pan2003,Gu2015PhD,Wang2019PhD}. These calculations provide insight into the RB-spectra as a function of relevant variables. In Fig.~\ref{Fig:Angle}(b) the dependence on the scattering angle $\theta$ is again displayed, now also for elevated pressures. While the overall width follows the pattern of the pure Rayleigh scattering, the effect of the Brillouin side peaks is apparent mainly in the forward scattering geometry, for small angles $\theta$, for the highest pressure and the lowest temperatures values. This implies that the Brillouin phenomenon and the testing of models can be most sensitively conducted in a forward scattering geometry. This forms the rationale of choosing a smaller scattering angle, in this study $\theta=55.7^\circ$, than in most other studies performed thusfar~\cite{Witschas2010,Witschas2011,Witschas2014,Gu2014b,Shang2019b}.

In Fig.~\ref{Fig:RB-Lineshape} RB-spectra calculated within this Tenti-S6 formalism are displayed, as a function of incident wavelength, temperature and pressure. The results show that for longer wavelengths, for lower temperatures, and for higher pressures, the Brillouin side peaks become more pronounced. The Brillouin side peaks and their distinctness are, in general terms, associated with the collisional and internal relaxation behaviour of gases. So in order to extract information on these relaxation phenomena occurring in gases, e.g. extracting information on the bulk viscosity $\eta_{\rm b}$, conditions must be chosen under which the RB-profiles undergo a decisive effect. The simulations show that such information can best be obtained from gases at low temperatures and at high pressures. As for the wavelength it would in principle be favorable to use a long wavelength, but a consideration is the strong decrease of the scattering cross section~\cite{Strutt1899,Sneep2005}, scaling with $\lambda^{-4}$, that deteriorates the signal-to-noise ratio in light scattering experiments. Moreover at longer wavelengths ($\lambda > 900$ nm) the quantum efficiency of photon detectors decreases severely. Here a wavelength of 532 nm is chosen since it is well detectable and an often used wavelength in atmospheric lidar experiments.

In Fig.~\ref{Fig:RB-Transport} simulations of the spectral lineshapes are plotted, for standard instrumental settings of $\theta = 90^{\circ}$ and $\lambda_i=532$ nm and for a pressure of 1 bar and room temperature. In the simulations, where a particle mass of $m=29$ amu is adopted for air, the values of the thermal conductivity $\kappa$, shear viscosity $\eta_{\rm s}$, and the internal heat capacity c$_{\rm int}$ as well as the ratio of bulk viscosity over shear viscosity $\eta_{\rm b}/\eta_{\rm s}$ are varied over a range around the known literature values.  These simulations show that the RB-spectral profiles only marginally depend on the gas transport coefficients. Even for large deviations of the parameters, only small shifts in the line profiles are found. Specifically for the case of the bulk viscosity a variation of 50\% in $\eta_{\rm b}$ causes a maximum variation of 2\% in the Brillouin side peaks.

The thermal conductivity and shear viscosity of air under the given standard conditions
are known to good accuracy according to the Sutherland formulas or Sutherland laws~\cite{White2006}:
\begin{equation}
\label{Eq:ShearViscosityCalc}
  \eta_{\rm s} = \eta_{\rm s}^{0} \left( \frac{T}{T_0} \right)^{3/2} \left( \frac{T_0+S_{\rm \eta}}{T+S_{\rm \eta}} \right)
\end{equation}
and
\begin{equation}
\label{Eq:ThermalCondCalc}
  \kappa = \kappa_{\rm 0} \left( \frac{T}{T_0} \right)^{3/2} \left( \frac{T_0+S_{\rm th}}{T+S_{\rm th}}\right) \ ,
\end{equation}
where ${T_0}$ = 273 K, $\eta_{\rm s}^{0} = 1.716 \times 10^{-5}$ Pa$\cdot$s, $\kappa_{\rm 0}$ = 0.0241 W/K$\cdot$m, $S_{\rm \eta}$ = 111 K, $S_{\rm th}$ = 194 K. This leads to a value for the thermal conductivity for air of
$\kappa$ = 2.55$\times 10^{-2}$ W/m$\cdot$K and a value for the shear viscosity of $\eta_{\rm s}$ = 1.83$\times10^{-5}$ Pa$\cdot$s, both at room temperature. For the bulk viscosity a value of $\eta_{\rm b}$ = 1.48$\times10^{-5}$ Pa$\cdot$s is adopted in the simulations~\cite{Gu2014b}.

As was noticed in previous studies on light scattering the values of some transport coefficients depend on the excitation frequency. This holds in particular for the bulk viscosity, which was found to deviate by some four orders of magnitude for the case of CO$_2$ between measurements performed at acoustic frequencies, i.e. in the MHz range, or light scattering experiments for GHz scattering frequencies~\cite{Pan2005,Gu2014a,Wang2019}.
Also for the case of the thermal conductivity such a frequency dependence was found, specifically for N$_2$O~\cite{Wang2018}.
This phenomenon is associated with relaxation of the internal degrees of freedom of the molecule, i.e., rotations and vibrations~\cite{Clark1972,Wang2019}.
When the product of the sound frequency and the relaxation time  $f_{\rm s}\tau_{\rm i(i = vib, rot)} \gg 1$,  the corresponding internal degree of freedom will become frozen.
Generally, the rotational relaxation time $\tau_{\rm rot}$ is about 10$^{-10}$ seconds, that is, $f_{\rm s}\tau_{\rm rot} \approx 1$ and the rotation will be excited. For N$_2$ and O$_2$, the vibrational relaxation time $\tau_{\rm vib}$ is larger than 10$^{-4}$ seconds at room temperature~\cite{Taylor2013}, which means the vibrational degrees remain frozen, both for acoustic (MHz) and light scattering frequencies (GHz). For air, which mainly includes N$_2$ and O$_2$, we suppose that the vibrational degree of freedom is not excited and the thermal conductivities are the same for low and high frequencies. Hence we adopt the values as given by the Sutherland relations and the values reported in Ref.~\cite{White2006}. These values are in agreement for other reported values for the shear viscosity and thermal conductivity of air~\cite{Lemmon2004,Montgomery1947}.

For completeness also the dependence of the RB-spectra profile on the internal heat capacity c$_{\rm int}$  is displayed in Fig.~\ref{Fig:RB-Transport}(c). For air, composed of diatomic molecules with two degrees of rotational relaxation, c$_{\rm int}$ = 2/2R, where R is the gas constant. A calculation is performed for a triatomic molecular species,  with c$_{\rm int}$ = 3/2R, and otherwise the same gas transport coefficients. The calculation shows a small dependency on molecular composition via its dependence on c$_{\rm int}$.

As illustrated in Figs.~\ref{Fig:Angle},~\ref{Fig:RB-Lineshape}, and \ref{Fig:RB-Transport} the RB-spectral line shapes exhibit more pronounced Brillouin side peaks for certain conditions where the effects of collisional and internal relaxation phenomena become apparent. For those conditions the information content of the measured spectra is enhanced, allowing for extraction of the gas transport coefficients related to collisional phenomena.
The scaling of the sharpness and the pronouncedness of the Brillouin side peaks, here calculated via the Tenti-model, can be approached in a more direct manner by considering that the width of the side peaks is physically connected to damping of elastic waves in the medium, as was analyzed by Leontovich~\cite{Leontovich1941}.
Based on a model for damping of elastic waves in a liquid medium the magnitude of the broadening of the Brillouin peak due to damping was derived (in units of angular frequency):
\begin{equation}\label{Eq:BrillouinLinewidth}
  \Gamma_B = \frac{\Omega_B^2}{v^2_s} \Gamma
\end{equation}
where $\Omega_B$ is the Brillouin shift as defined in Eq.~(\ref{Bril-shift}) and
$\Gamma$ a damping parameter in the Navier-Stokes equation, characterizing the classical dissipation of the acoustic mode~\cite{Fabelinskii2012,Rossing2014}:
\begin{equation}
  \Gamma = \frac{1}{\rho_0}\{\frac{4}{3}\eta_{\rm s} + \eta_{\rm b}+ \frac{\kappa}{c_p}(\gamma -1)\} \ ,
   \label{Eq:fabel-2}
\end{equation}
with $\rho_0$ the density of the medium,  $\gamma = c_p/c_v$, and the other thermodynamic parameters as defined before.
This approach leads to a ratio of the Brillouin shift $\Omega_B$ over the Brillouin linewidth $\Gamma_B$ representing sharpness and the degree to which the Brillouin side peaks are resolved \cite{Fabelinskii2012}:
\begin{equation}
  \frac{ \Omega_B}{\Gamma_B} = \frac{v_s}{2n k_i \Gamma}\frac{1}{\sin \theta/2}
  \label{fabel-1}
\end{equation}
Even though this model derives its validity from hydrodynamics in essence it will be applicable to RB-scattering in the kinetic regime, thus providing an explanation and a scaling law for the separation of Brillouin side peaks at forward scattering angles $\theta$. In principle the side peaks are most pronounced at the smallest value of $\theta$, but as Fabilinskii already discussed, a further experimental limitation lies in the instrument width of the measurement apparatus~\cite{Fabelinskii2012}. That should be sufficiently small to resolve the Brillouin side peaks at gradually smaller shifts $\Omega_B$ when approaching forward scattering.

Note that the dependence of the RB-spectral lineshape on the measurement conditions ($\theta$, $\lambda_i$, $T$ and $p$) is much more pronounced than for the gas transport coefficients ($\eta_{\rm s}$, $\kappa$ and $\eta_{\rm b}$).
In particular the bulk viscosity $\eta_{\rm b}$, cannot be determined easily from gas-transport experiments and its values are not well known, while Fig.~\ref{Fig:RB-Transport}(d) shows that the RB-profile is only marginally dependent on the exact structure of the Brillouin side lobes in the scattering spectrum. This implies that spectra of high quality are needed to extract a reasonably accurate value of the bulk viscosity, while at the same time the measurement conditions have to be set or measured at high accuracy.

Concluding from these simulations a strategy can be formulated to measure $\eta_{\rm b}$ at conditions of pronounced Brillouin side bands, and then extrapolate its value to those conditions where the side bands are less pronounced, possibly taking into account the scaling of $\eta_{\rm b}$ as a function of values for some of the other transport coefficients. That is the rationale for a measurement of RB-spectra at a scattering angle of $\theta=55.7^\circ$ than scattering at $\theta=90^\circ$, cf. Fig.~\ref{Fig:Angle}(b).

\begin{figure*}
  \centering
  \includegraphics[width=0.95\linewidth]{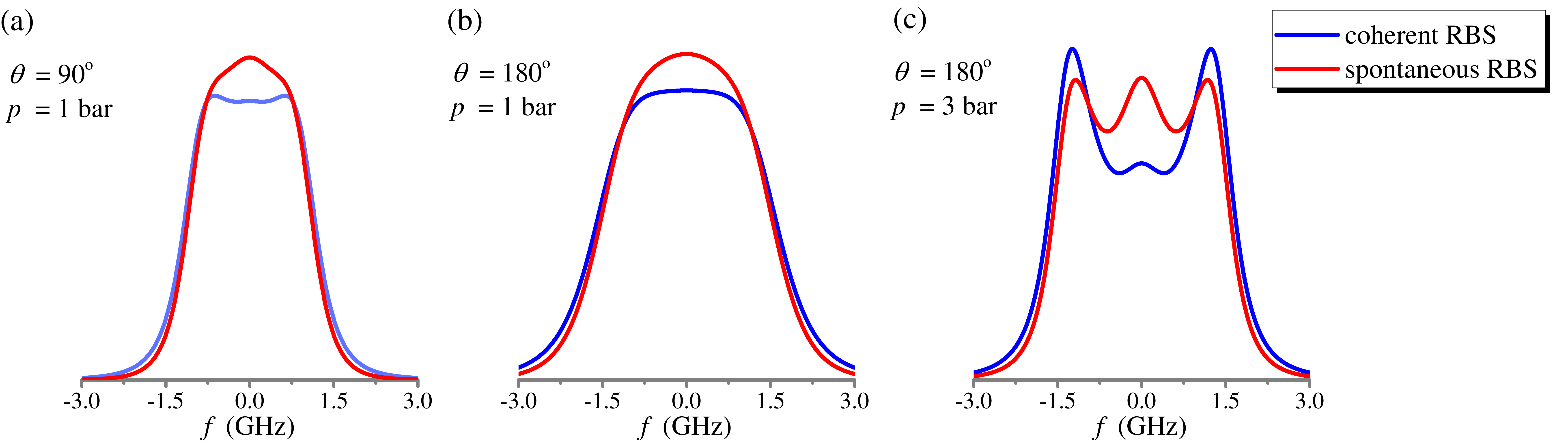}
  \caption{Simulated spectral line shapes for coherent RBS and spontaneous RBS in air by the Tenti-S6 model.
  Conditions are $T=293.15$ K and $\lambda_i=532$ nm and nominal values of the transport coefficients for air. Pressures $p$ and scattering angles $\theta$ as indicated.}
  \label{Fig:CRBSandSRBS}
\end{figure*}

\vspace{0.5cm}

The present study focuses on \emph{spontaneous} Rayleigh-Brillouin scattering (RBS), and most of the experimental studies on RBS have employed this form of light scattering.
However, through the availability of narrowband pulsed lasers \emph{coherent} RBS has been explored as well~\cite{She1983,Grinstead2000,Pan2002,Pan2002b,Pan2004,Vieitez2010,Meijer2010,Manteghi2011,Graul2014}. In such studies an experimental configuration is setup where laser beams drive the density fluctuations. Dipole forces, proportional to the molecular polarizability $\alpha$, and induced by the pump beams (with amplitudes $E_1$ and $E_2$) create a standing-wave lattice at wave vector $\boldsymbol k$ = $\boldsymbol {k_1}-\boldsymbol {k_2}$ ($\boldsymbol{k_i}$ the wave vectors of the pump fields). This effect produces an acceleration:
\begin{equation}
    a(x,t) = - \frac{ \alpha k E_1 E_2}{2m} \sin(kx-\omega t) \ ,
\end{equation}
which adds as an additional term~\cite{Pan2004} to the kinetic WCU equation, which then reads:
\begin{eqnarray}
  \left(  \frac{\partial }{\partial t} + \boldsymbol v \cdot \nabla  + \boldsymbol a \cdot \nabla_v \right)
    f_i(\boldsymbol r,\boldsymbol v,t) = \\ \nonumber
    \sum_{jgl} \int |\boldsymbol v- \boldsymbol v_1| \sigma^{gl}_{ij} (f_g'f_l' - f_if_j) d\Omega d^3v_1 \ .
    \label{Eq:FullWCUFuncCRBS}
\end{eqnarray}

These driven, coherent, density fluctuations add to the thermal fluctuations probed in spontaneous RBS, and give rise to a different line profile under conditions of coherent RBS. The line profiles can be calculated via an extended version of the Tenti-S6 model, which was developed by Pan and coworkers~\cite{Pan2003,Pan2004}. Via this formalism RB-spectral profiles are calculated for typical conditions of air ($p = 1$ bar, $T=293.15$ K), for a typical wavelength $\lambda_i=532$ nm used in cases of pulsed lasers, and for scattering angles $\theta =90^\circ$ and $180^\circ$, while the transport coefficients are set at the nominal values for these conditions. In Fig.~\ref{Fig:CRBSandSRBS} a comparison is made between line shapes for coherent RBS and spontaneous RBS for the same conditions, including the same scattering angle $\theta$. The predicted spectra show that the Brillouin side peaks become more pronounced in the case of coherent RBS. So in principle the underlying collisional and relaxation parameters may be measured at better accuracy compared to spontaneous RBS. However, it is noted that typical experimental configurations for coherent RBS employ a geometry with a scattering angle of close to $180^\circ$; in fact $\theta=178^\circ$ is used in many studies~\cite{Grinstead2000,Pan2002,Meijer2010,Manteghi2011}. Under such angles distinctiveness of the side-peaks deteriorates.

\section{Experimental} \label{sec:ExperimentalGreen}

\begin{figure}
  \centering
  \includegraphics[width=0.5\linewidth]{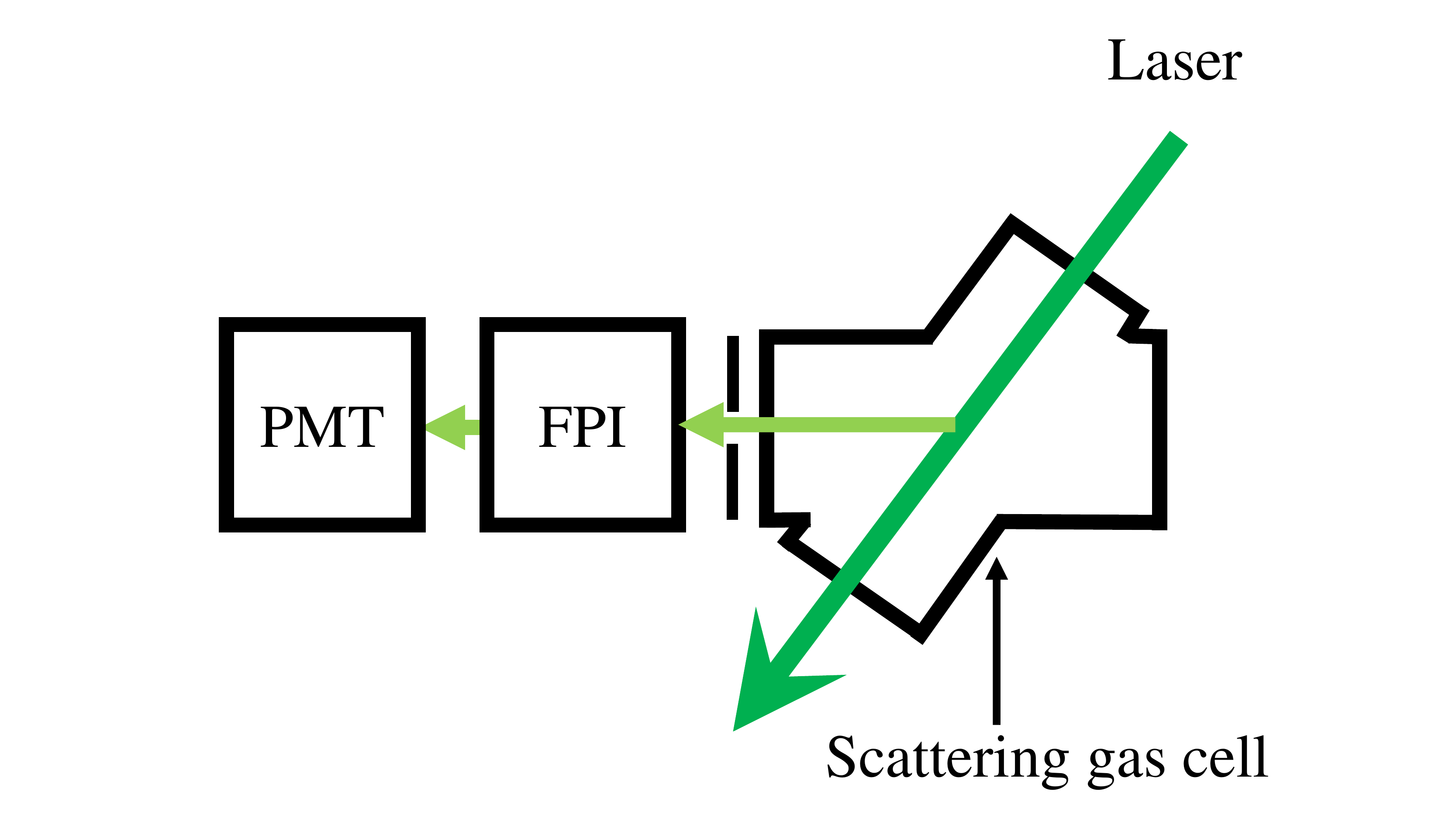}
  \caption{Schematic sketch of the experimental setup. The green light (532.22 nm, full green line) propagates through the scattering gas cell and produces the scattered light. The scattered light at scattering angle $\theta = 55.7^\circ$ is analyzed by a FPI (Fabry-Perot interferometer) and detected on a PMT (photomultiplier tube).}
  \label{Fig:GreenSetupSimple}
\end{figure}

Rayleigh-Brillouin scattering spectra of air are measured in an experimental setup, which was used in a previous study~\cite{Wang2019} and is schematically shown in Fig.~\ref{Fig:GreenSetupSimple}. The laser light with wavelength of 532.22 nm, power of 5 Watts, and bandwidth less than 5 MHz travels through a gas cell. For this scattering cell, two Brewster-angled windows are mounted at entrance and exit ports to reduce the loss of incident light and the inside walls were painted black to reduce stray light. A pressure gauge is connected to the cell to monitor the pressure change and a temperature control system consisting of PT-100 sensor, Peltier elements as well as water cooling are used to keep the cell at a constant temperature with uncertainty less than $0.1 \: ^\circ{\rm C}$.

Data were recorded in a pressure regime of 0.25 - 3 bar and for temperatures of 273.2 - 333.2 K.
The forward RB-scattering was recorded under an angle $\theta = 55.7 \pm 0.3^\circ$, measured by a home-built goniometer rotation stage. The opening angle is less than $\Delta\theta = 0.5^\circ$ determined by the geometry of the scattering gas cell and a slit positioned behind the gas cell. The scattered light propagates through a bandpass filter (Materion, T$>$ 90$\%$ at $\lambda_i$ = 532 nm, bandwidth $\triangle\lambda$ = 2.0 nm) onto a half-confocal Fabry-Perot interferometer (FPI).

The FPI is aligned using a reference laser beam. The radius of the curved mirror is $r = -12.5$ mm with reflectivity of 99\%. The FPI has a nominal free spectral range (FSR) of some 12 GHz, but this turns into an effective FSR if the 4-mode transmission pattern in considered~\cite{Gu2012rsi}. The precise value of the FSR is determined through frequency-scanning a laser (a narrowband tunable cw-ring dye laser) over more than 1000 modes of the FPI, while measuring the laser wavelength by a wavelength meter (Toptica HighFinesse WSU-30), and yields an uncertainty in the FSR below 1 MHz and an FSR=$2.9964$ GHz.

The instrument width, yielding a value of $\sigma_{\nu_{\rm instr}}$ = 58.0 $\pm$ 3.0 MHz (FWHM), is determined by using the reference beam while scanning the piezo-actuated FPI, following methods discussed by \citet{Gu2012rsi}
The instrument function is verified to exhibit a functional form of an Airy function, which may be well approximated by a Lorentzian function during data analysis.

RB-scattering spectral profiles were recorded by piezo-scanning the
FPI at integration times of 1 s for each step, usually over 18 MHz. A full spectrum covering a large number of consecutive RB-peaks and 10,000 data points were obtained in about 3 h. The piezo-voltage scans were linearized and converted to a frequency scale by fitting the RB-peak separations to the calibrated FSR-value~\cite{Gu2012rsi,Wang2019PhD}.

\section{Results}

\begin{figure}
  \centering
  \includegraphics[width=0.5\linewidth]{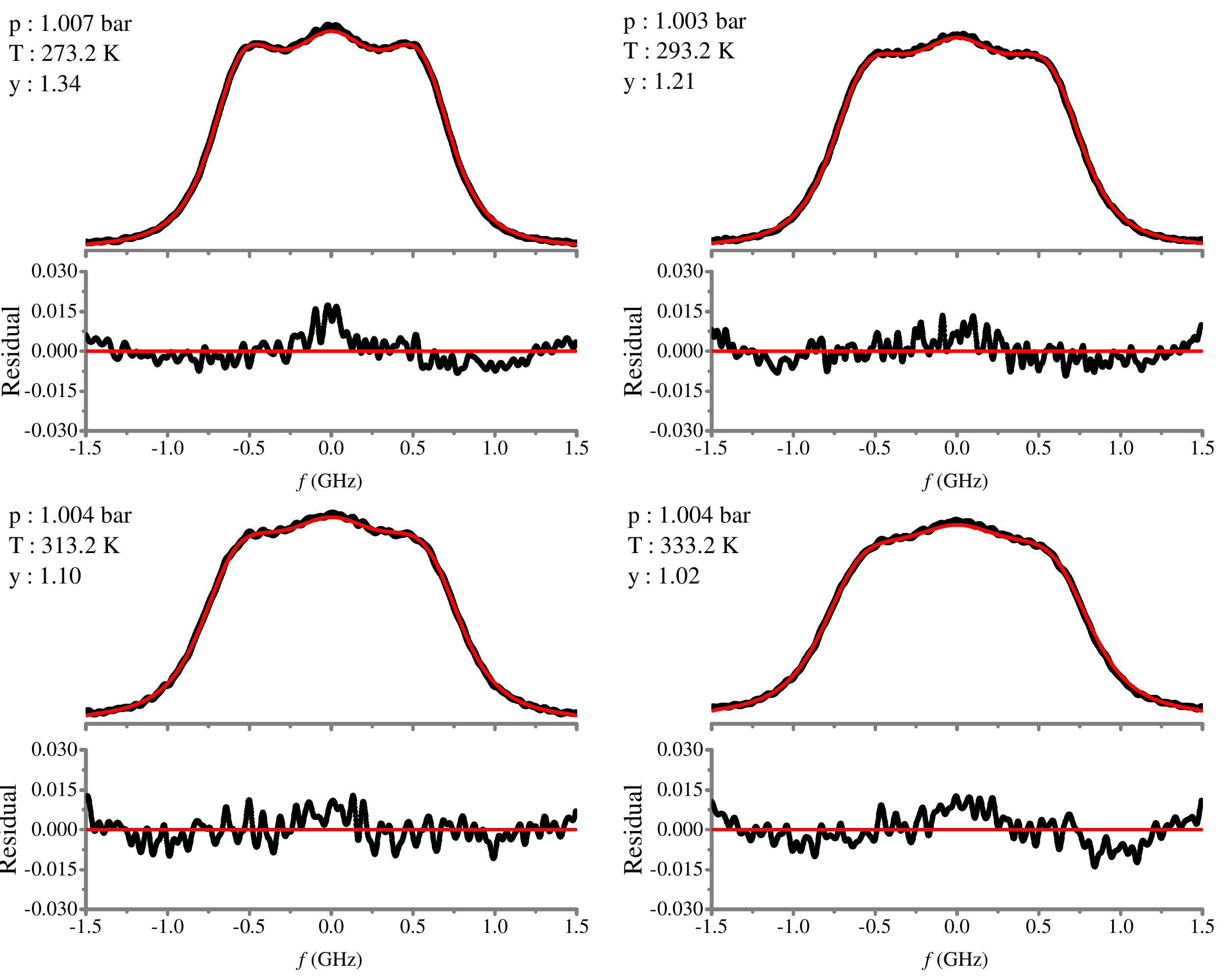}
  \caption{Experimental Rayleigh-Brillouin scattering profiles (black) of air at $p=1$ bar and $T=273.2 - 333.2$ K, and comparison with optimized Tenti-S6 model (red). Bottom graphs display the corresponding residuals. The experimental data were measured at wavelength of $\lambda_i$ = 532.22 nm and scattering angle of $\theta$ = 55.7$^\circ$, and these spectra are on a scale of normalized integrated intensity.}
  \label{Fig:Air532nm1bar}
\end{figure}

\begin{figure}
  \centering
  \includegraphics[width=0.5\linewidth]{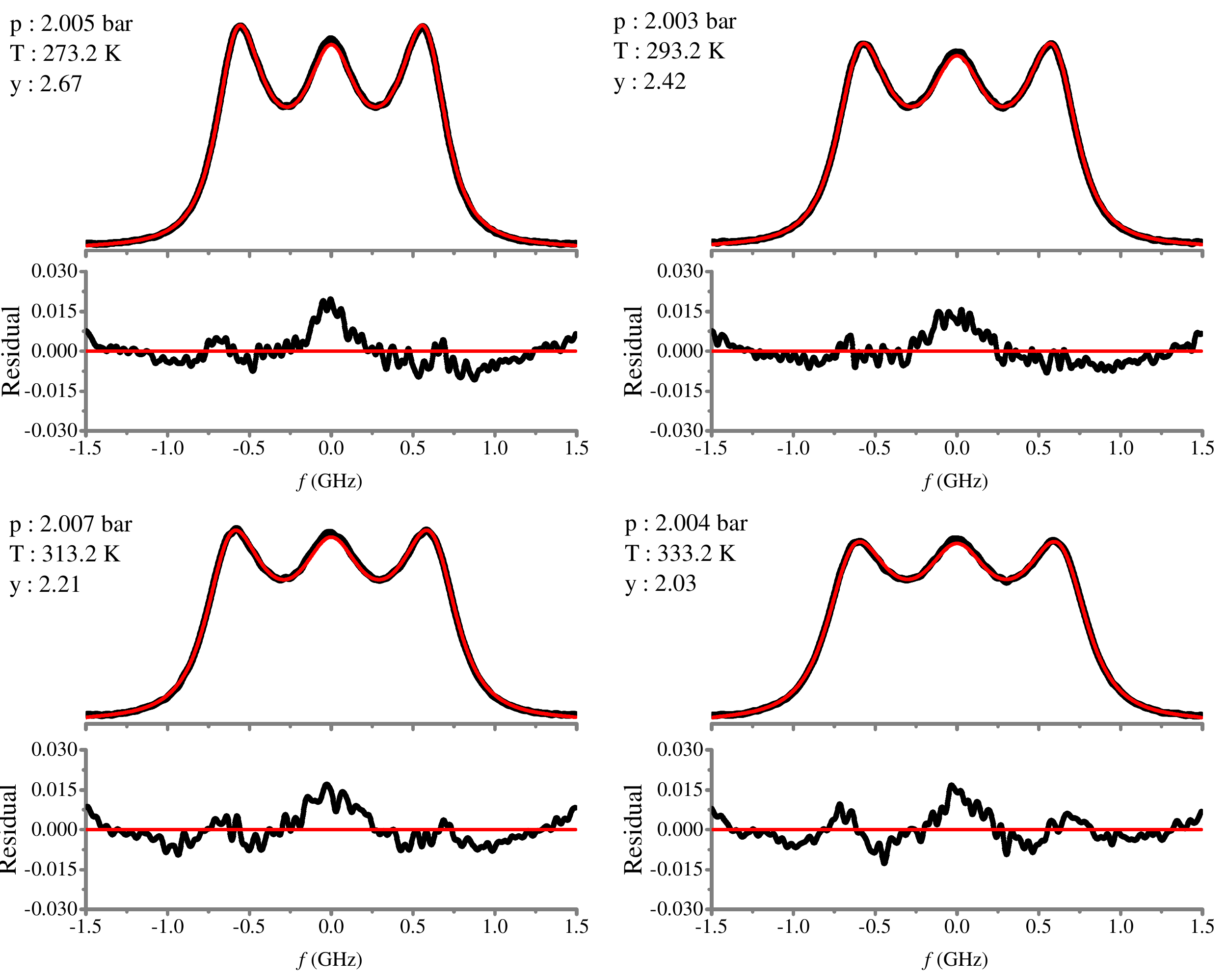}
  \caption{Experimental Rayleigh-Brillouin scattering profiles (black) of air at $p=2$ bar and comparison with optimized Tenti-S6 model (red). Further details as in Fig.~\ref{Fig:Air532nm1bar}.}
  \label{Fig:Air532nm2bar}
\end{figure}

\begin{figure}
  \centering
  \includegraphics[width=0.5\linewidth]{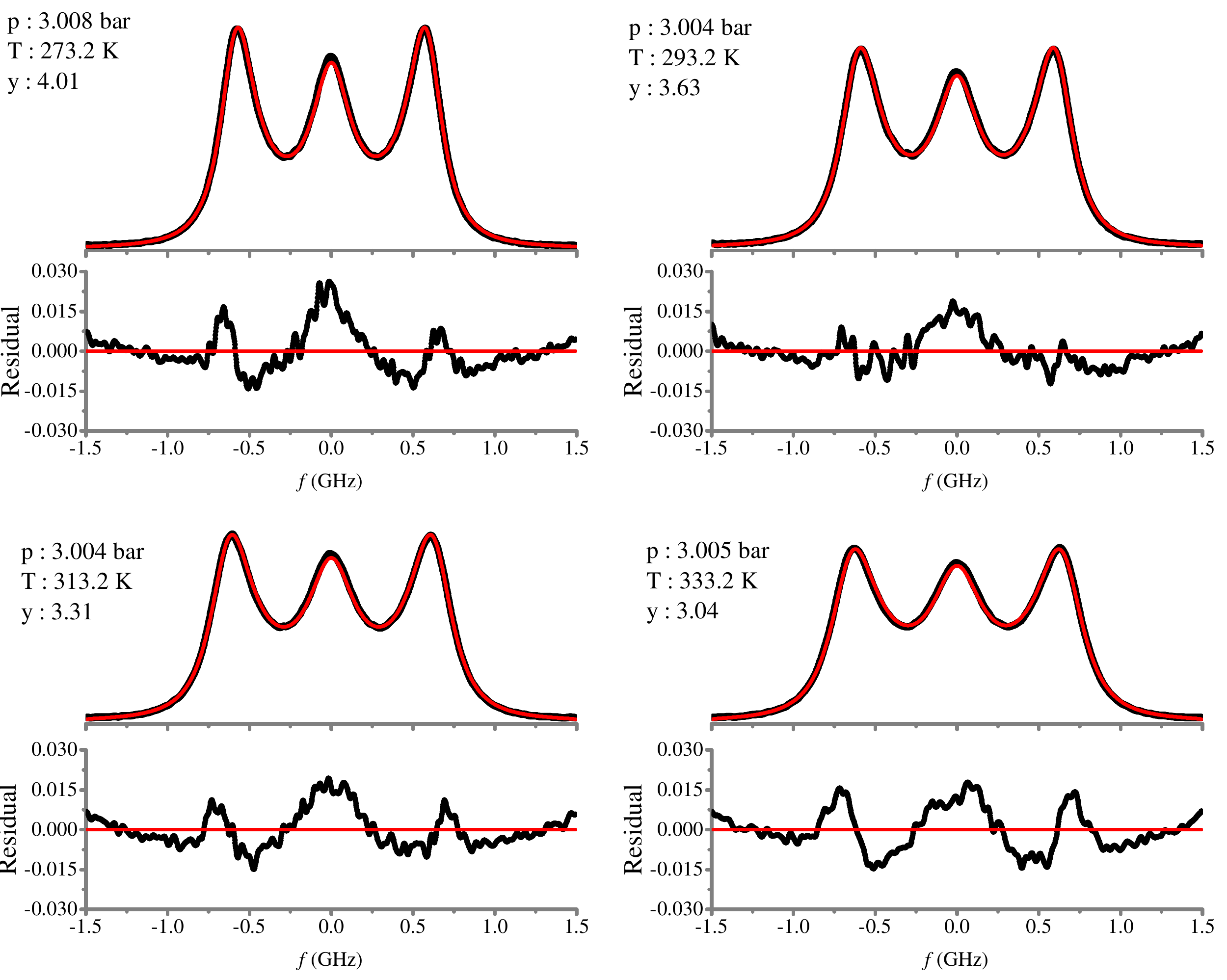}
  \caption{Experimental Rayleigh-Brillouin scattering profiles (black) of air at $p=3$ bar and comparison with optimized Tenti-S6 model (red). Further details as in Fig.~\ref{Fig:Air532nm1bar}.}
  \label{Fig:Air532nm3bar}
\end{figure}

The experimental data of spontaneous RBS spectra of air were measured at a wavelength of $\lambda_i=532.22$ nm, a scattering angle of $\theta = 55.7 \pm 0.3^\circ$ and at various pressures and temperatures. The resulting spectra are split into two subsets, one set obtained at the higher pressures $p=1 - 3$ bar, for deducing and verifying the values of the gas-transport coefficients, in particular to extract a value for the bulk viscosity $\eta_{\rm{b}}$, based on the Tenti-S6 model. The data for these measurements are displayed in Figs.~\ref{Fig:Air532nm1bar}, \ref{Fig:Air532nm2bar} and \ref{Fig:Air532nm3bar}.
A second subset, for data obtained at sub-atmospheric pressure conditions, $p=0.25-0.75$ bar, are then used to verify the obtained model description in the realm of atmospheric applications, to be discussed in Section~\ref{Atmos}. The conditions under which all data were obtained are listed in Table~\ref{Tab:GreenAir}.

\begin{table}
  \tbl{Determination of bulk viscosities $\eta_{\rm b}$ from RBS measurements on air at $\lambda=532.22$ nm and ($p,T$) experimental conditions, corresponding to uniformity parameters $y$ as indicated.}
  {\begin{tabular}{c c c c }
  \hline
  \hline
    $p$ (bar) & $T$(K) & $y$ & $\eta_{\rm b}$\, ($10^{-5}$ Pa$\cdot$s)\\
  \hline
    0.255& 273.2 &  0.34  \\
    0.254& 293.2 &  0.31   \\
    0.254& 313.2 &  0.28 \\
    0.253& 333.2 &  0.26\\
  \hline
   0.505 & 273.2 &  0.67    \\
   0.504 & 293.2 &  0.61 \\
   0.504 & 313.2 &  0.55  \\
   0.504 & 333.2 &  0.51\\
  \hline
   0.754 & 273.2 &  1.01   \\
   0.755 & 293.2 &  0.91 \\
   0.754 & 313.2 &  0.83 \\
   0.756 & 333.2 &  0.76\\
  \hline
    1.007 & 273.2   & 1.34  & 1.23\,(0.03)  \\
    1.003 & 293.2   & 1.21  & 1.55\,(0.12)\\
    1.004& 313.2    & 1.10  & 1.80\,(0.23) \\
    1.004& 333.2    & 1.02  & 2.19\,(0.19)\\
  \hline
   2.005 & 273.2   &  2.67  & 1.03\,(0.04) \\
   2.003 & 293.2   & 2.42   & 1.16\,(0.03)\\
   2.007 & 313.2   & 2.21   & 1.27\,(0.03) \\
   2.004 & 333.2   & 2.03   & 1.45\,(0.04)\\
  \hline
  3.008  & 273.2   & 4.01   & 0.93\,(0.04)\\
  3.004  & 293.2   & 3.63   & 1.08\,(0.05)\\
  3.004  & 313.2   & 3.31   & 1.16\,(0.03)\\
  3.005  & 333.2   & 3.04   & 1.36\,(0.04)\\
  \hline
  \hline
  \end{tabular}}
  \label{Tab:GreenAir}
\end{table}

\begin{figure}
  \centering
  \includegraphics[width=0.5\linewidth]{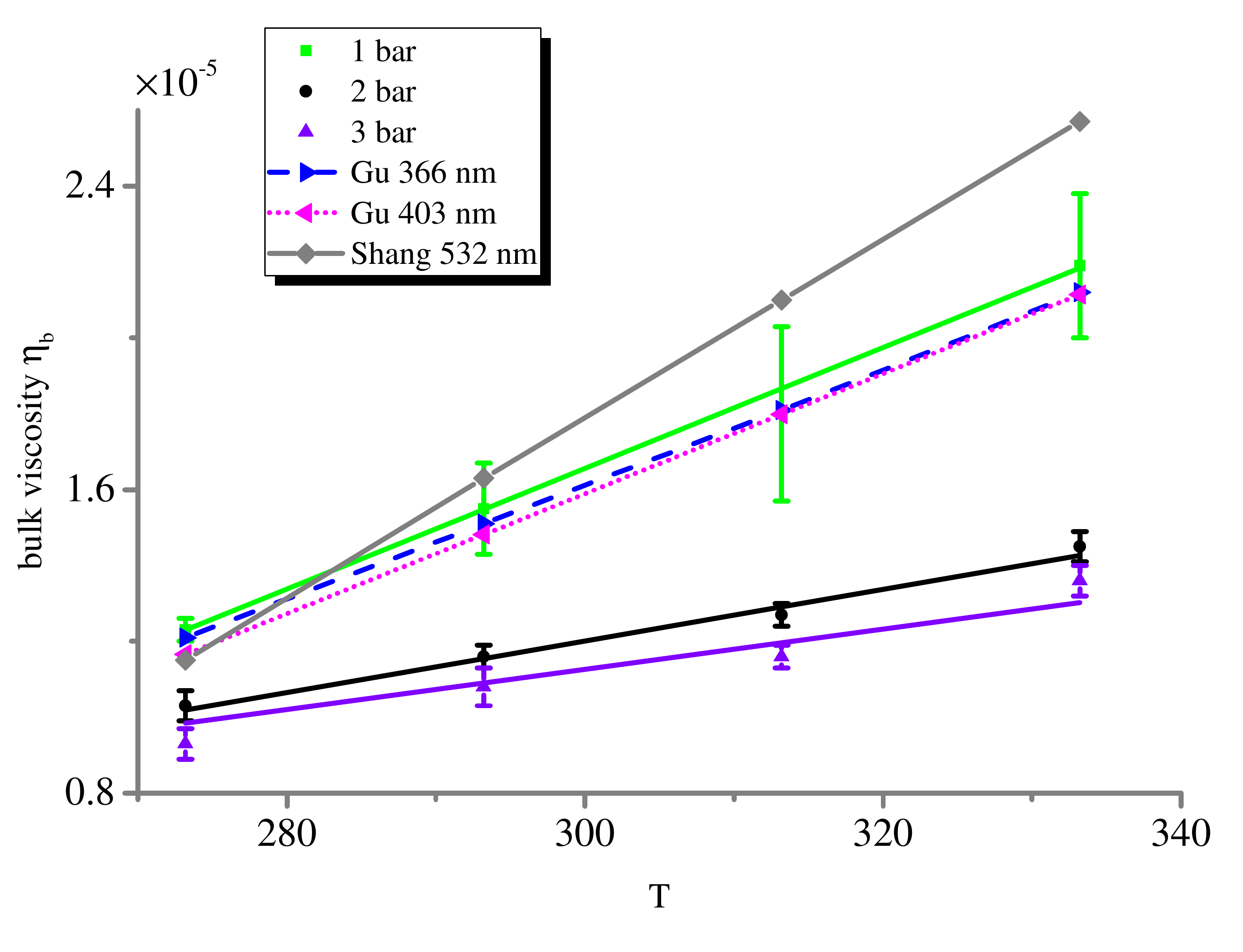}
  \caption{Bulk viscosities of air derived from RB-spectra at pressures of 1 bar (green square), 2 bar (dark circle), 3 bar (purple triangle). The lines represent the values calculated from Eq.~\eqref{Eq:BulkCalc} and the constants $\eta^T_{\rm b}$ as listed in Table~\ref{Tab:eta_b_p_T}. A comparison is made with results from \citet{Gu2013a} at $\lambda_i=366$ nm (blue), and  $\lambda_i=403$ nm (magenta), as well as \citet{Shang2019b} (gray) for $\lambda_i=532$ nm.}
  \label{Fig:BulkviscosityCompare}
\end{figure}

The data of the first subset were incorporated in a Tenti-S6 model description, for which a Matlab-code was written~\cite{Wang2019PhD}. Here, the values for the shear viscosity $\eta_{\rm s}$ and thermal conductivity $\kappa$ were taken for the specific measurement temperatures, interpolated from the Sutherland formulas given in Eqs.~\eqref{Eq:ShearViscosityCalc} and \eqref{Eq:ThermalCondCalc}. A fit was made to extract an optimum value for the bulk viscosity $\eta_b$ via a least-squares procedure
for each ($p$, $T$) combination and plotted in Fig.~\ref{Fig:BulkviscosityCompare} as a function of temperature. Also data from  previous investigations are included~\cite{Gu2013a,Shang2019b}.

Inspection of the deduced values of  $\eta_{\rm b}$ indicates that all values fall within the range $1.0 - 2.0 \times 10^{-5}$ Pa$\cdot$s, with some pressure dependence, although not systematic. However, the temperature dependencies follow an approximately linear temperature dependence for all individual data sets. To describe this a functional form is chosen to represent this behaviour:
\begin{equation}
\label{Eq:BulkCalc}
  \eta_{\rm b} = \eta_{\rm b}^{0}  + \eta_{\rm b}^{T} \cdot (T - T^{\rm{ref}}) \ ,
\end{equation}
where the reference temperature $T^{\rm{ref}}$ = 250 K is chosen such that the results for various pressures, and also for various previous experiments exhibit a common origin point, at $\eta_{\rm b}^{0}$ = 0.86\,(0.01)$\times 10^{-5}$ Pa$\cdot$s.
The resulting parameters representing the monotonous increase of the bulk viscosity versus temperature $\eta_{\rm b}^{T}$ are listed in Table~\ref{Tab:eta_b_p_T}, while the slopes are included in Fig.~\ref{Fig:BulkviscosityCompare}.

\begin{table}
  \centering
\caption{Values of $\eta_{\rm b}^{T}$.} \label{Tab:eta_b_p_T}
  \begin{tabular}{c c}
  \hline
  \hline
  $p$ (bar) & $\eta_{\rm b}^{T}$ ($\times 10^{-7}$ Pa$\cdot$s/K) \\
  \hline
  1& 1.59\,(0.02) \\
  2& 0.68\,(0.02) \\
  3& 0.53\,(0.04) \\
  \hline
  \hline
  \end{tabular}
\end{table}

In Fig.~\ref{Fig:BulkviscosityCompare} also the resulting values for $\eta_{\rm b}$ from previous studies on RBS in air, all measured at a scattering angle of $\theta= 90^\circ$, are included~\cite{Gu2013a,Gu2014b,Shang2019b}.
The result of Fig.~\ref{Fig:BulkviscosityCompare} shows the spread in results for the bulk viscosity obtained between the present study and three previous studies~\cite{Gu2013a,Gu2014b,Shang2019b}
The simulations performed in Section~\ref{simulations} and displayed in Figs.~\ref{Fig:RB-Lineshape} and \ref{Fig:RB-Transport} show how insensitive the RBS line shape in fact is for the exact value of the bulk viscosity $\eta_{\rm b}$, and its much greater sensitivity to experimental parameters as the scattering angle $\theta$, and the readings of pressure and temperature. As for the temperature reading it may be noted that for all present and previous results a rather smooth temperature functionality is found, indicating that there should be no source of error connected to temperature. In order to deduce a set of values for the bulk viscosity of air an average is taken over the concatenated data sets. Since all data follow the behavior of Eq.~\eqref{Eq:BulkCalc}, crossing at a common point at $\eta^0_{\rm b}$ the entirety of results can be represented by a single value of $\eta^T_{\rm b}$. From a weighted average over all data a mean value of $\eta^T_{\rm b}= 1.29\,(0.23) \times 10^{-7}$ Pa$\cdot$s/K results,
which can be considered as a generic value for the calculation of RBS profiles in air, in conjunction with the other gas transport coefficients.

\section{Atmospheric RB spectral profiles}
\label{Atmos}

In the previous sections a description is given, based on the Tenti-S6 model on how to compute Rayleigh-Brillouin spectra profiles of air under various conditions and settings of an experiment. The scattering profile depends on some experimental settings as the scattering angle $\theta$ and the incident wavelength $\lambda_i$, as well as on some conditions of the gas, such as temperature and pressure, where air is treated as a hypothetical molecule with mass $m=29.0$ amu and $c_{\rm int}=(2/2)R$. Furthermore the RBS line shape depends on the thermodynamic gas properties of air, the shear viscosity $\eta_{\rm s}$ and the thermal conductivity $\kappa$, values of which can be determined from generic measurements on the gas~\cite{White2006,Montgomery1947,Lemmon2004}. Here it must be considered that these gas transport coefficients may be temperature dependent, as modeled in the Sutherland formulas, and that in most cases the vibrational relaxation is frozen in air under regular atmospheric conditions. Finally, the last thermodynamic gas property of air, the bulk viscosity can be described by $\eta_{\rm b} = 0.86 \times 10^{-5} + 1.29 \times 10^{-7}\cdot(T - 250)$.  This combined set of parameters, in combination with the Tenti-S6 model, should provide a universal description for RB-scattering in air, under all experimental settings and gas conditions, the latter with restriction to the kinetic regime.

\begin{figure}
  \centering
  \includegraphics[width=0.5\linewidth]{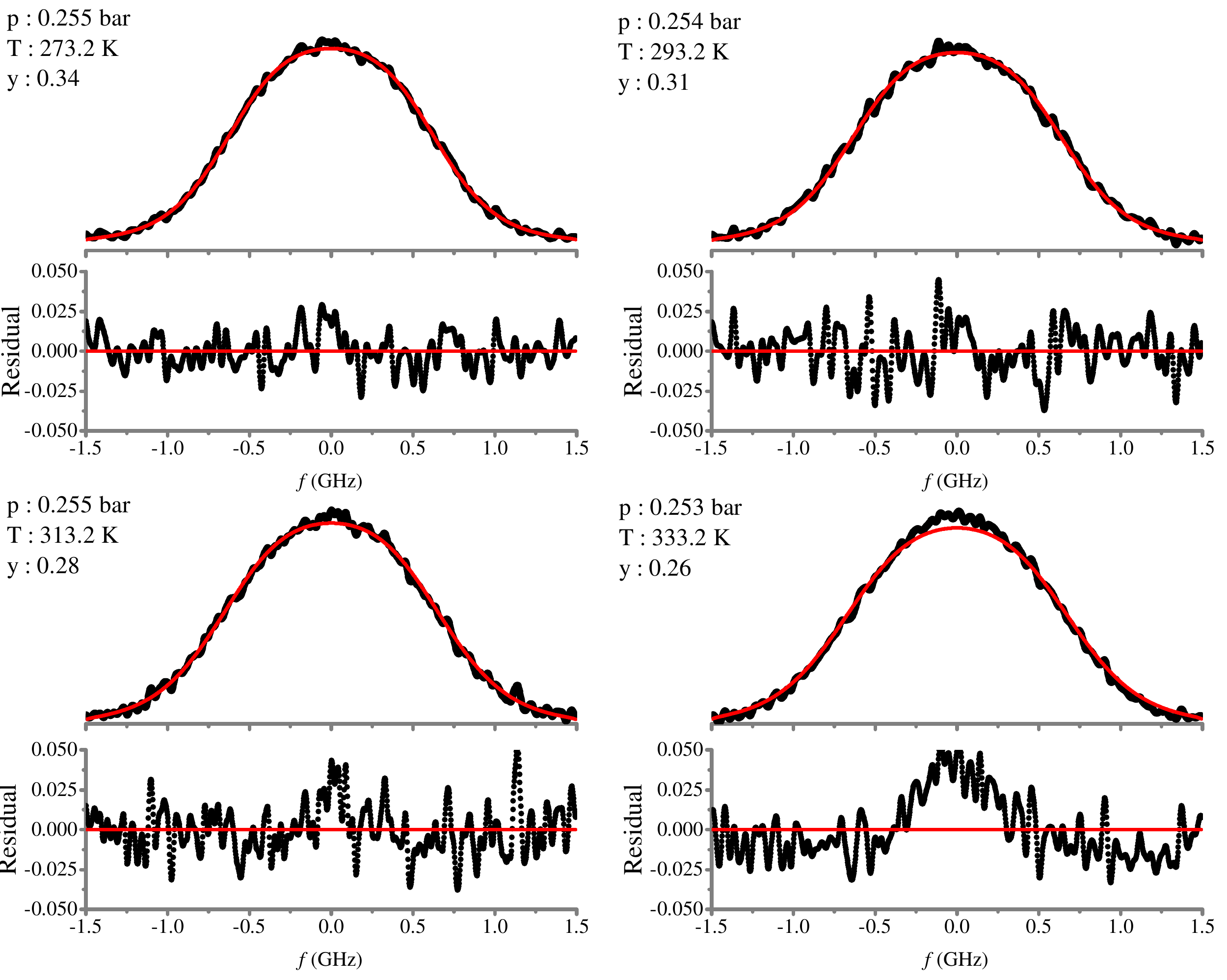}
  \caption{Experimental Rayleigh-Brillouin scattering profiles (black) of air at $p=0.25$ bar and temperature of 273.2 - 333.2 K, and comparison with Tenti-S6 model (red), based on fixed gas transport coefficients.  Bottom graphs display the corresponding residuals. The experimental data were measured at wavelength of $\lambda_i$ = 532.22 nm and scattering angle of $\theta$ = 55.7$^\circ$.}
  \label{Fig:Air532nm0.25bar}
\end{figure}

\begin{figure}
  \centering
  \includegraphics[width=0.5\linewidth]{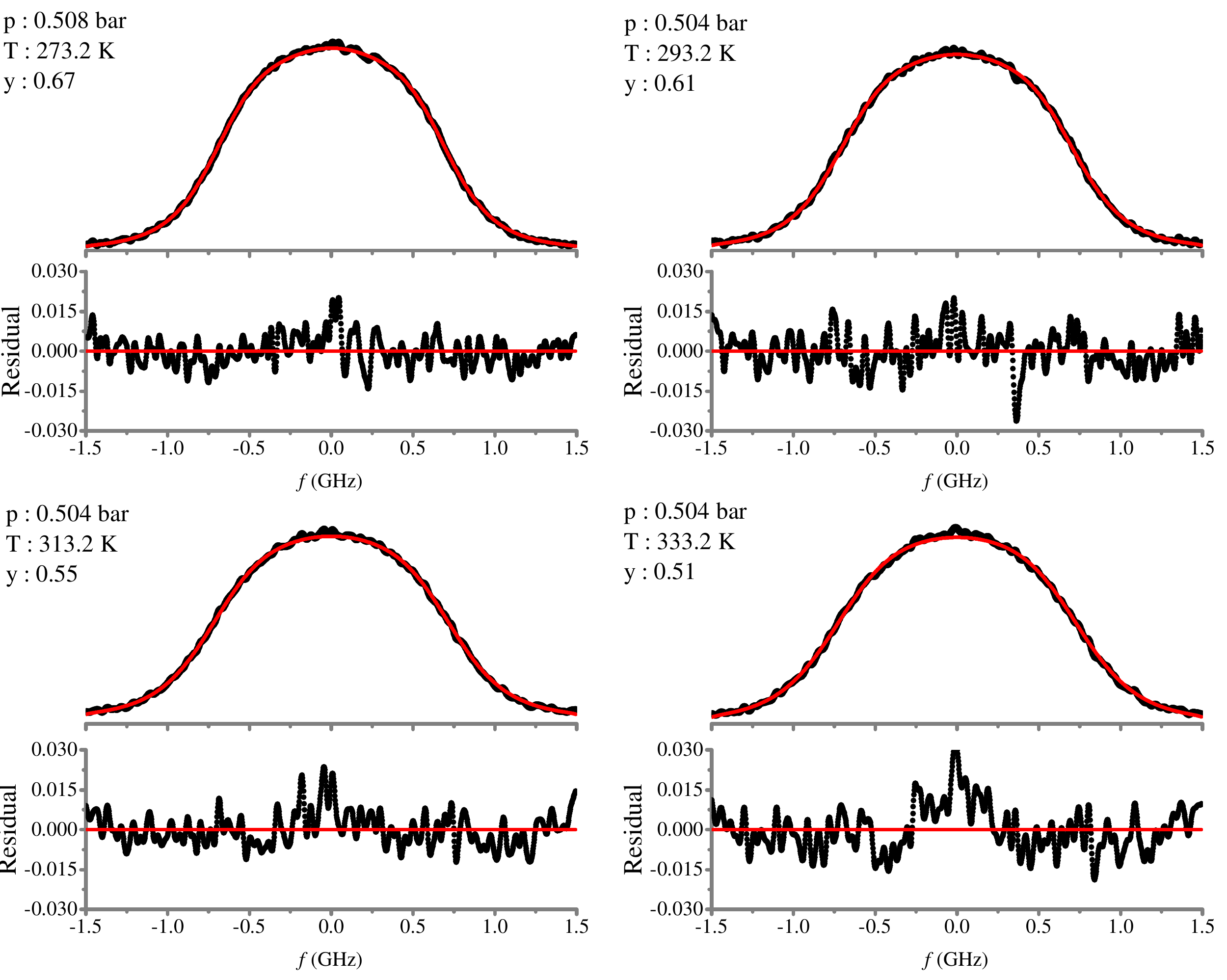}
  \caption{Experimental Rayleigh-Brillouin scattering profiles (black) of air at $p=0.5$ bar, and comparison with Tenti-S6 model (red), based on fixed gas transport coefficients.}
  \label{Fig:Air532nm0.5bar}
\end{figure}

\begin{figure}
  \centering
  \includegraphics[width=0.5\linewidth]{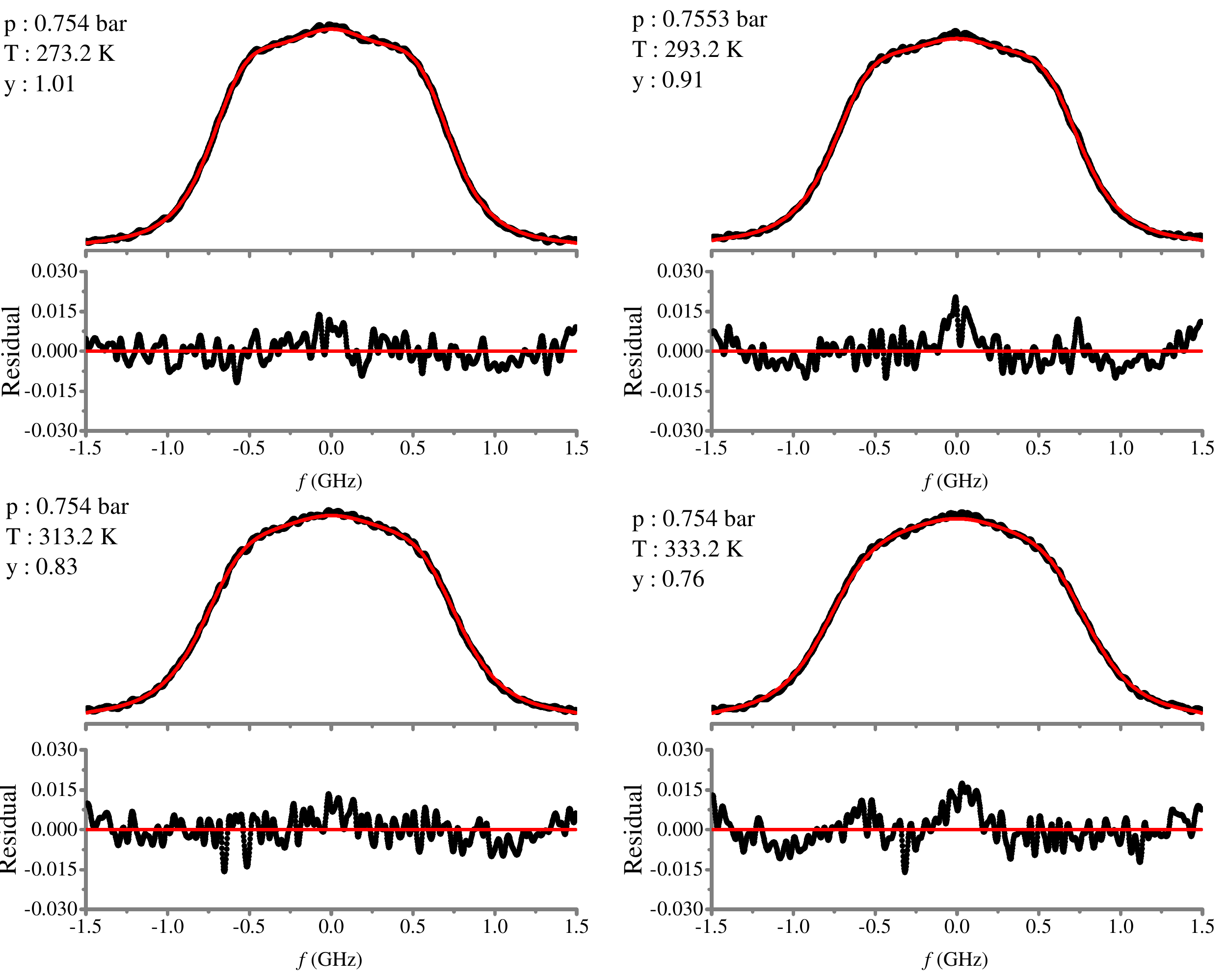}
  \caption{Experimental Rayleigh-Brillouin scattering profiles (black) of air at $p=0.75$ bar, and comparison with Tenti-S6 model (red), based on fixed gas transport coefficients.}
  \label{Fig:Air532nm0.75bar}
\end{figure}

This model description is now used to compute the RB-spectral profiles under conditions measured in the present experiment for low pressures.
Experimental and computed spectra for pressures of 0.25 bar, 0.50 bar and 0.75 bar, and in all cases for a variety of temperatures in the range 273 - 333 K, are compared in Figs.~\ref{Fig:Air532nm0.25bar},~\ref{Fig:Air532nm0.5bar}, and~\ref{Fig:Air532nm0.75bar}. The spectra show that good agreement is found, well within an error margin of 1\% of the peak intensity. In a few cases a small residual in the form of a positive photon signal is found exactly at the center frequency. This phenomenon was observed in some of our previous studies~\cite{Witschas2014}, and can be safely ascribed to spurious Mie reflection of dust particles. Hence, it may be concluded that the set of parameters finds very good agreement with the observed RB-scattering profiles of air measured under atmospheric conditions of $p < 1$ bar.

\section{Universal scaling of RB-scattering in air}

As is shown in the above the Tenti-S6 model provides a representation of the RB-light scattering function based on the conditions of the scattering process and the specific conditions of the gas. The Tenti-S6 model provides also a means to transfer the scattering profile from one set of conditions to another, a transfer based on the knowledge of the relevant underlying parameters.
Such scaling can be connected to the concept of uniformity in kinetic gas theory, where  universal scaling parameters are represented in dimensionless coordinates ($y$,$x$) to describe the light scattering response of the gas.
The dimensionless uniformity parameter $y$ is proportional to the ratio of scattering wavelength $\Lambda$ (connected to the scattering wave vector as $\Lambda=2\pi/q$) to the mean-free-path $L_{\rm{mfp}}$ between collisions:
\begin{equation}
y=\frac{1}{2\pi} \frac{\Lambda}{L_{\rm{mpf}}} \ .
\end{equation}
Note that in the community of rarefied gas dynamics the rarefaction parameter $\delta_{\rm rp}$ is also used, proportional to the uniformity parameter via $\delta_{\rm rp} = 2\pi y$.
With $L_{\rm{mfp}}=v_0/\alpha_{col}$, $v_0$ the thermal velocity of the molecules, and the collisional frequency $\alpha_{col}=p/\eta_s$, $p$ and $\eta_{\rm s}$ being the pressure and the shear viscosity of the gas, the uniformity parameter can be written as:
\begin{equation}
\label{Eq:Parameter-y}
    y= \frac{\lambda_i/n}{4\pi \sin{\theta/2}} \frac{p}{\eta_{\rm s}\sqrt{2k_BT/m}} \ .
\end{equation}
It may be verified (see Figs.~\ref{Fig:Angle} - \ref{Fig:RB-Transport}) that the isolated prominence and the distinctiveness of the Brillouin side peaks is highest for the largest values of the uniformity parameter $y$. Large $y$ corresponds with higher pressure $p$, lower temperature $T$, smaller scattering angle $\theta$ and longer wavelength $\lambda_i$.

Also the frequency scale of the RB scattering profiles can be converted to a reduced and dimensionless frequency scale defined as
\begin{equation}
\label{Eq:Parameter-x}
  x  = \frac{\omega}{q v_0 } \ ,
\end{equation}
where $q$ is the scattering vector and $\omega = 2 \pi f$, with $f$ the frequency of the incident light beam.

This description in terms of uniformity in a rarefied gas dynamics approach considers the gas as independent of the gas transport coefficients,
which relate to internal relaxation mechanisms, and to the intermolecular dynamics connected to a potential describing the interactions. In this context the shear viscosity $\eta_{\rm s}$ is a parameter connected to the kinetics describing the molecular collisions in terms of a mean-free-path. Such scaling in terms of dimensionless parameters was discussed and tested for gases with atomic constituents~\cite{Fookson1976,Ghaem1980a,Ghaem1980b}. The scaling is expected to hold for densities for which the Boltzmann equation may be applied, thus where the mean field or average interaction between molecules may be neglected.

\begin{figure}
  \centering
  \includegraphics[width=0.5\linewidth]{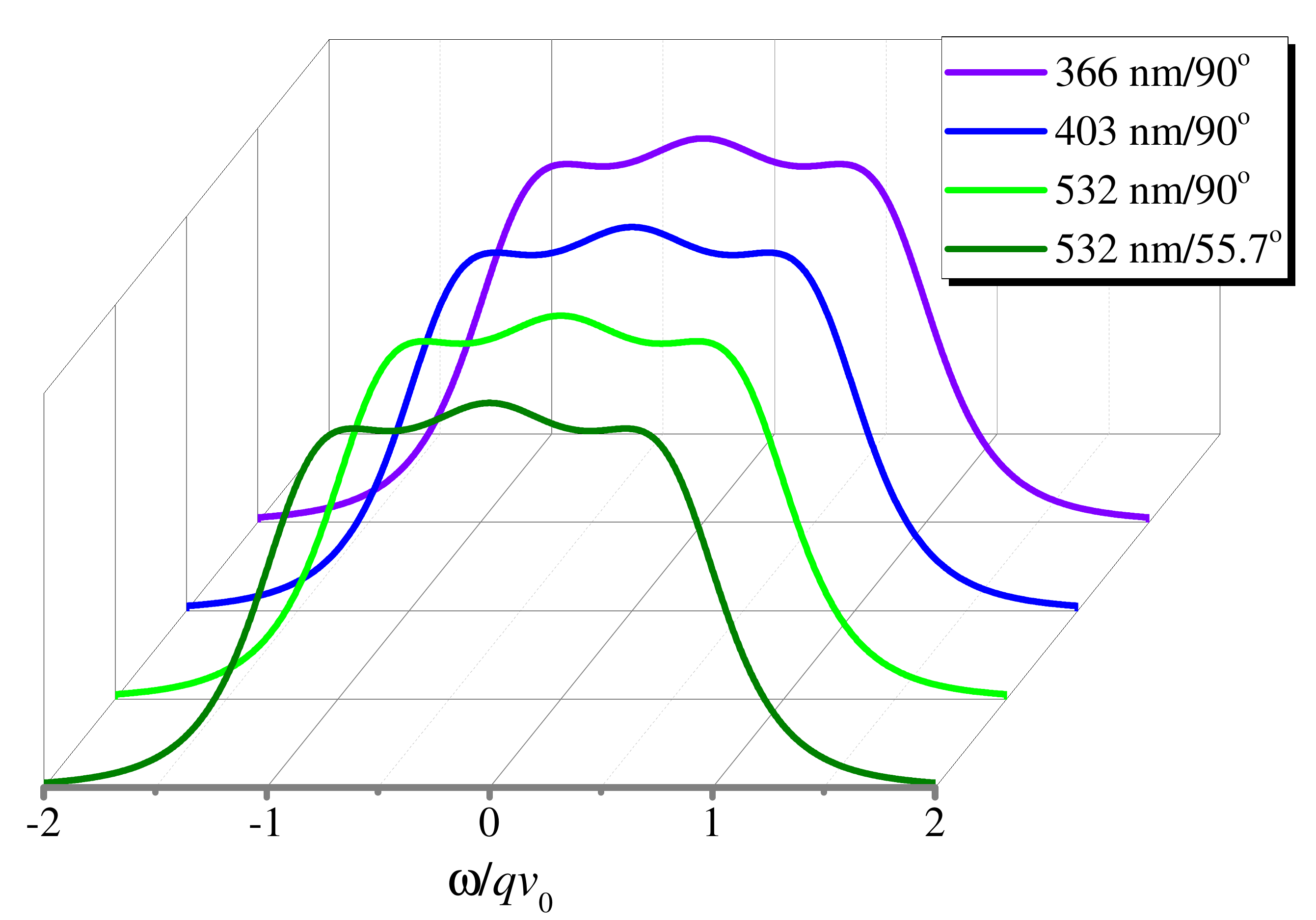}
  \caption{Simulated spectral line shapes for RB-scattering in air by the Tenti-S6 model for a uniformity parameter of $y = 1.10$, produced from different conditions (1) $\lambda_i=366.8$ nm, $\theta=90^\circ$; (2) $\lambda_i=403$ nm, $\theta=90^\circ$; (3) $\lambda_i=532$ nm, $\theta=90^\circ$; (4) $\lambda_i=532$ nm, $\theta=55.7^\circ$. Transport coefficients of $\kappa$ = 2.55$\times 10^{-2}$ W/m$\cdot$K, $\eta_{\rm s}$ = 1.83$\times10^{-5}$ Pa$\cdot$s, $\eta_{\rm b}$ = 1.48$\times10^{-5}$ Pa$\cdot$s are fixed.}
  \label{Fig:XscaleForSameY}
\end{figure}

Even though the  description of the light scattering process and its resulting spectral profile can be connected to dimensionless ($y$,$x$) coordinates, it is still needed to exploit an underlying model for generating spectra and for demonstrating the scaling of spectra. Such procedure is here undertaken with application of the Tenti-S6 model, where it must be considered that the Tenti-model involves aspects of gas transport. To avoid the effects of transport coefficients on the RBS-line shape their values are taken at fixed nominal values ($\kappa = 2.55\times10^ {-2}$ W/m$\cdot$K, $\eta_{\rm s} = 1.83\times10^ {-5} $ Pa$\cdot$s, and $\eta_{\rm b} = 1.48\times10^ {-5}$ Pa$\cdot$s), even though these coefficients may be temperature dependent, or be subject to freezing out of internal relaxation modes. Fixing the values of these thermodynamic properties implies that they do not affect the dimensionsless or universal scaling.

The outcome of such scaling is shown in Fig.~\ref{Fig:XscaleForSameY}, where calculated spectra are displayed for a single value of the dimensionless uniformity parameter $y=1.10$. Note that these RB-scattering spectra, plotted on a common dimensionless $x$-axis, are calculated for different combinations of
$\lambda_i$, $\theta$, $p$ and $T$.
While wavelengths (366 nm, 403 nm, and 532 nm) and scattering angles (90$^\circ$, 55.7$^\circ$) are taken to coincide with those used in experiments, values for $p$ and $T$ are chosen such as to match  $y=1.10$. In these calculations no convolution with an instrument function is included.
The results plotted in Fig.~\ref{Fig:XscaleForSameY} for four different experimental conditions indeed show identical RB-scattering profiles for a common uniformity parameter. For a pair of dimensionless ($y$,$x$) values the RB-scattering spectra are not influenced by the incident wavelength and scattering angle, nor of a set of $p$ and $T$. Hence, the RB-spectral can be expressed as a function of $y$ and $x$ alone. This demonstration of dimensionless scaling of calculated spectra verifies that spectra, recorded in one set of conditions can be transferred to another set of conditions.

This universal scaling was investigated and tested for experimental RB-spectra of noble gases~\cite{Fookson1976,Ghaem1980a,Ghaem1980b}.
The scaling in those studies was done for constant scattering conditions ($\theta$, $\lambda_i$), while combinations of gaseous settings ($p$, $T$) were chosen as to match a similar value of $y$.  In these studies spectra for different noble gases He, Ne and Ar were compared, which made it necessary to invoke a slight rescaling of the reduced frequency axis for the Brillouin shift $\omega_{\rm B}$.
As a result of the experimental comparisons it was concluded that the RB-spectra in the kinetic regime grossly follow ideal gas conditions and that they are rather insensitive to details of the intermolecular potential. However, small deviations were found, in particular at higher densities with gradual increase of deviations at larger values of $y$, that were attributed to variations in the sound velocity and thermal conductivity~\cite{Ghaem1980a}.

In the present study we extend this comparison of dimensionless scaling laws for RB-scattering to the case of air. In this specific case the difference between atoms and molecules needs to be considered, where molecules undergo internal relaxation between the modes of translation, vibration and rotation, which is expressed in terms of a bulk viscosity parameter $\eta_{\rm b}$. Experimentally the present data set obtained at $\lambda_i=532.22$ nm and $\theta=55.7^\circ$, and various values of $p$ and $T$, leading to uniformity parameters $y$ as listed in Table~\ref{Tab:GreenAir}, are compared with previously obtained data for RB-scattering in air at $\lambda_i=366.8$ nm \cite{Gu2013a} and $\lambda_i=403.00$ nm \cite{Gu2014b}. Hence the scaling comparison covers all ($\lambda_i$, $\theta$, $p$, $T$) parameters.

\begin{figure}
  \centering
  \includegraphics[width=0.5\linewidth]{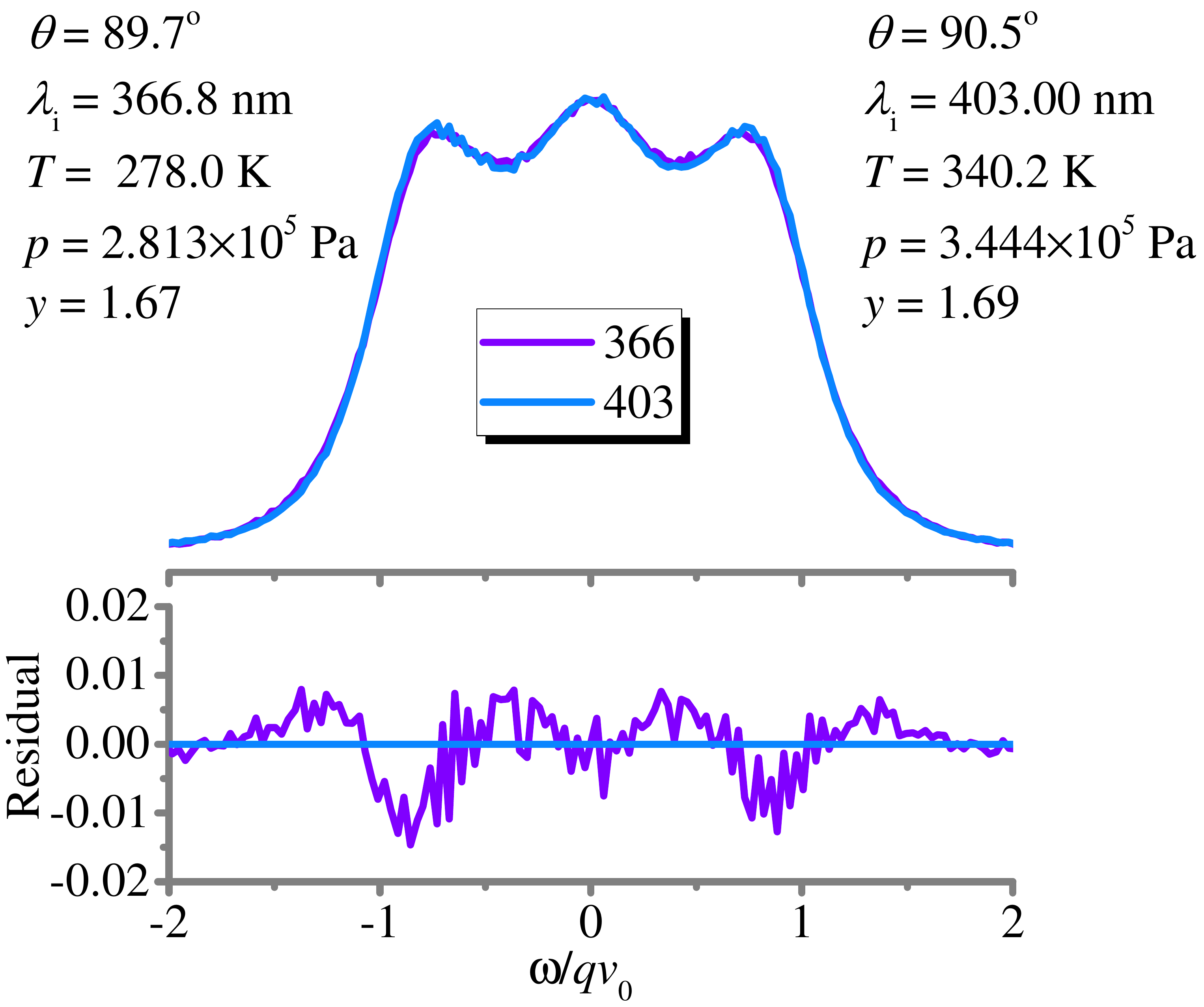}
  \caption{Comparison of experimental RB-spectra of air for a nearly coincident uniformity parameter $y=1.67 - 1.69$, measured at 366.8 nm and 403 nm and other conditions as indicated. }
  \label{Fig:Example1}
\end{figure}

\begin{figure}
  \centering
  \includegraphics[width=0.5\linewidth]{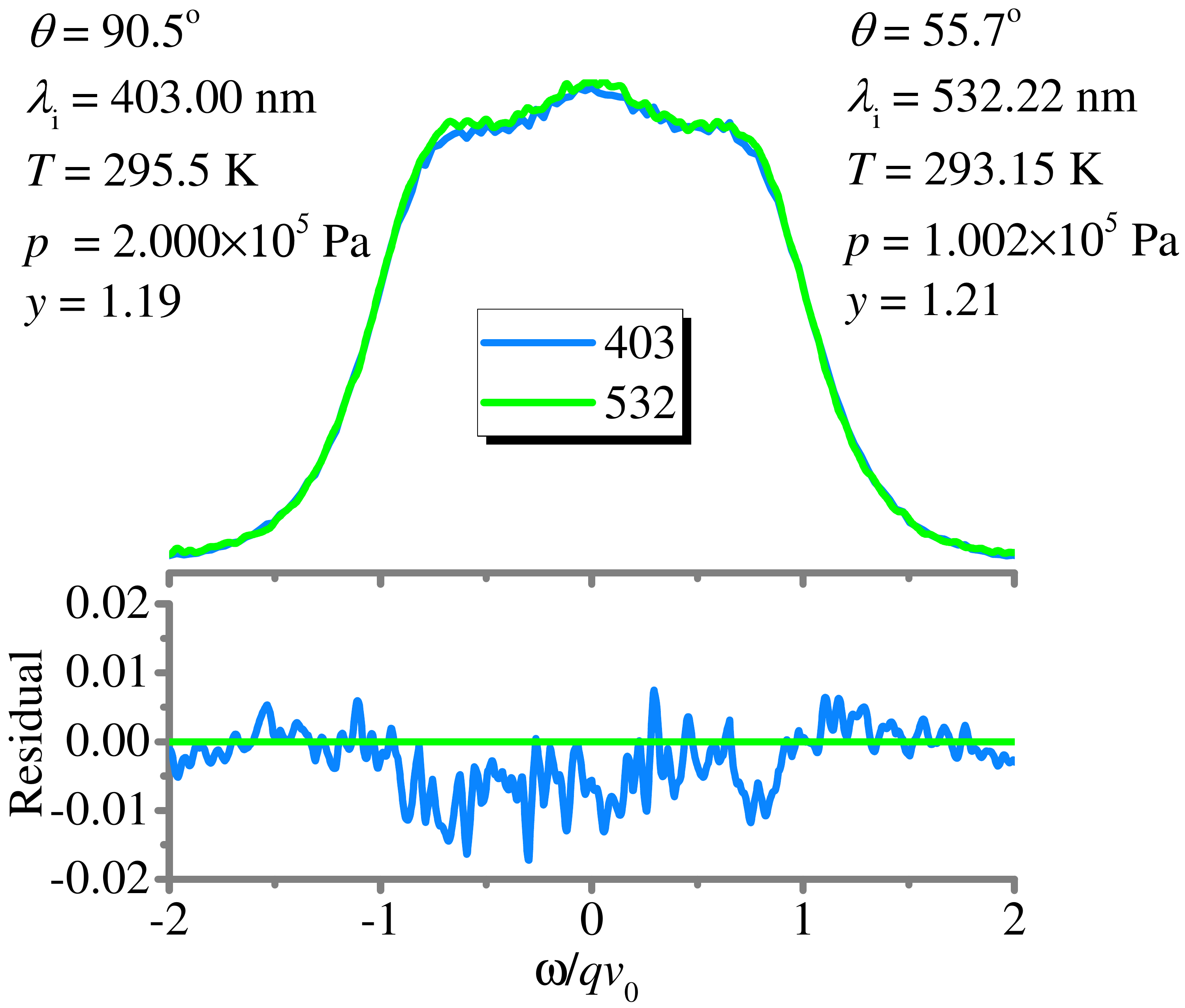}
  \caption{Comparison of experimental RB-spectra of air for a nearly coincident uniformity parameter $y=1.19 - 1.21$, measured at 403.00 nm and 532.22 nm and other conditions as indicated.}
  \label{Fig:Example2}
\end{figure}

The data sets of RB-spectra contain examples where the uniformity parameters $y$ are near-coincident, while the underlying physical conditions are vastly different, thus allowing for an experimental comparison of scaling, while remaining in the kinetic regime of $y \approx 1$. In Fig.~\ref{Fig:Example1} a comparison is made between spectra recorded for $\lambda_i=366.8$ and $\lambda_i=403.00$ nm, and other parameters as indicated, resulting in a uniformity parameter of $y= 1.67 - 1.69$.
In Fig.~\ref{Fig:Example2} a comparison is made for two cases also for a near-coincident uniformity parameter $y=1.19 - 1.21$ for wavelengths $\lambda_i=403.00$ nm and $\lambda_i=532.22$ nm.

The comparisons of Figs.~\ref{Fig:Example1} and \ref{Fig:Example2} demonstrate that the dimensionless scaling of the spectra at the two values of $y$ produces excellent agreement.  The residuals indicate agreement better than 1.5\% for the maximum differences in both cases, where it should be noted that the $y$-parameters do not match exactly, in both cases differing by 0.02.
Moreover the instrument widths used in the different experiments were different. When converted to the reduced $x$-axis scale the instrument widths were $\delta x _{({\rm 366 nm})}= 0.147$, $\delta x _{({\rm 403 nm})}= 0.097$, and for the present experimental study $\delta x _{({\rm 532 nm})}= 0.080$.

The aptness of the universal scaling approach for the case of air demonstrates that under the conditions of the included experiments, mainly set by the range of the uniformity parameter $y \approx 1$, air very much behaves as an ideal gas. Real gas effects as a result of molecular interactions, relaxation phenomena, freezing out of certain modes, and represented in terms of a bulk viscosity parameters, which may be dependent on temperature, only marginally affect the scaling and the shape of the Rayleigh-Brillouin spectra. Indeed, from the a priori simulations shown in Figs.~\ref{Fig:Angle} - \ref{Fig:RB-Transport} it is evident that the effect of small changes at the few \%-level in the bulk viscosity or thermal conductivity is only marginal and not detectable.

\section{Conclusion}

We have investigated Rayleigh-Brillouin scattering in air over a range of pressures and temperatures, measuring high quality data at a signal-to-noise ratio well within the 1\% level of peak intensities. Based on an a priori modeling a smaller scattering angle ($\theta=55.7^\circ$) was chosen for the experimental geometry, because under such condition the Brillouin side peaks become more pronounced. From the higher pressure data $p \geq 1$ bar values of the bulk viscosity parameter were derived. From these measurements, and including information from previous studies, a temperature dependent functional form is established for this elusive gas transport coefficient $\eta_{\rm b}$.

Based on the Tenti-S6 model description and a set of reliable gas transport coefficients for air a universal framework is established to predict Rayleigh-Brillouin scattering profiles under various density and temperature conditions of atmospheric air, and for different settings of the scattering geometry and incident wavelength. This framework holds for various scattering geometries and experimental parameters, such as the incident wavelength. The available data thus are at the basis of a predictive tool for delivering RB-spectra of air, for application in atmospheric sensing.

Finally from the now available experimental data sets for RB-scattering a universal scaling is tested relating spectra to a dimensionless uniformity parameter and a reduced frequency scale. Very good agreement is found for the regime of $y \approx 1$, demonstrating that in the range of atmospheric pressures and temperatures air very much behaves like an ideal gas.

\section*{Supplementary Material}
The Supplementary Material consists of three parts. The Tenti-model is described in detail, while the Matlab code for calculating Rayleigh-Brioullin spectra via the Tenti-S6 model is provided. In addition the data files of the recorded RBS spectra for the various ($p$, $T$) conditions are included.

\section*{Acknowledgement}
This research was supported by the China Exchange Program jointly run by the Netherlands Royal Academy of Sciences (KNAW) and the Chinese Ministry of Education. YW acknowledges support from the Chinese
Scholarship Council (CSC) for his stay at VU Amsterdam. The authors wish to thank Willem van de Water (Delft University) for fruitful discussions.


\end{document}